\documentclass[journal]{IEEEtran}
\usepackage{amsmath,epsfig,color}
\usepackage{url}
\usepackage{bigstrut,multirow,rotating}
\usepackage{textcomp,booktabs}
\usepackage{tabularx}
\usepackage{floatrow}
\newfloatcommand{capbtabbox}{table}[][\FBwidth]
\usepackage{colortbl}
\usepackage{xcolor}
\usepackage{multirow}
\usepackage{booktabs}
\usepackage{underscore}
\usepackage[marginal]{footmisc}
\usepackage{subfigure}
\usepackage{algorithm,algorithmic}
\usepackage{amssymb}
\usepackage{graphicx}
\usepackage{bbding}
\usepackage{makecell}
\usepackage{cite}

\begin{document}
\newcommand{\bl}[1]{{\color{black}#1}}
\newcommand{\bbl}[1]{{\color{red}#1}}
\title{ No-Reference Image Quality Assessment by Hallucinating Pristine Features}
\author{Baoliang Chen, Lingyu Zhu, Chenqi Kong, Hanwei Zhu, Shiqi Wang,~\IEEEmembership{Senior Member,~IEEE} and Zhu Li,~\IEEEmembership{Senior Member,~IEEE}

\thanks{This work is supported  in part by the National Natural Science Foundation of China under 62022002, in part by Shenzhen Virtural University Park, The Science Technology and Innovation Committee of Shenzhen Municipality (Project No: 2021Szvup128), in part by the Hong Kong Research Grants Council General Research Fund (GRF) under Grant 11203220.}

\thanks{B. Chen, L. Zhu, C. Kong, H. Zhu and S. Wang are with the Department of Computer Science, City University of Hong Kong, Hong Kong (e-mail: blchen6-c@my.cityu.edu.hk; lingyzhu-c@my.cityu.edu.hk; cqkong2-c@my.cityu.edu.hk; hwzhu4-c@my.cityu.edu.hk; 
shiqwang@cityu.edu.hk). Z. Li is with the Department of Computer Science and Electrical Engineering, University of Missouri-Kansas City, MO 64110, USA. (email:zhu.li@ieee.org). Corresponding author: Shiqi Wang. 
}

}

%\markboth{Submitted to IEEE Transactions on Image Processing}%
\maketitle

%------------------------------------------------------------------------------------------
\begin{abstract}

In this paper, we propose a no-reference (NR) image quality assessment (IQA) method via feature level pseudo-reference (PR) hallucination. \bl{The proposed quality assessment framework is rooted in the view that the perceptually meaningful features could be well exploited to characterize the visual quality, and the natural image statistical behaviors are exploited in an effort to deliver the accurate predictions.}
%The proposed quality assessment framework is grounded on the prior models of  natural image statistical behaviors, and rooted in the view that the perceptually meaningful features could be well exploited to characterize the visual quality. 
%The  Instead of predicting the reference in an image-level, we learn the reference information in the feature-level, getting rid of the design of a specific network for PR image generation. 
Herein, the PR features from the distorted images are learned by a mutual learning scheme with the pristine reference as the supervision, and the discriminative characteristics of PR features are further ensured with the triplet constraints. Given a distorted image for quality inference, the feature level disentanglement is performed  with an invertible neural layer for final quality prediction, leading to the PR and the corresponding distortion features for comparison. 
%Due to the quality estimated by our model are patch-wised, a gated recurrent unit (GRU)  based quality aggregation module is proposed to aggregate the predicted quality scores of different patches in an adaptive manner. 
The effectiveness of our proposed method is demonstrated on four popular IQA databases, and superior performance on cross-database evaluation also reveals the high generalization capability of our method. The implementation of our method is publicly available on
\textit{https://github.com/Baoliang93/FPR}.
\end{abstract}

\IEEEpeerreviewmaketitle

%------------------------------------------------------------------------------------------
\section{Introduction}

\IEEEPARstart{I}{mage} quality assessment (IQA), which aims to establish the quantitative connection between the input image and the corresponding perceptual quality, serves as a key component in a wide range of computer vision applications \cite{guo2017building,liu2017quality,zhang2017learning}. The typical full-reference (FR) IQA models resort to fidelity measurement in predicting image quality via measuring the deviation from its pristine-quality counterpart (reference). 
The pioneering studies date back to 1970’s and a series of visual fidelity measures have been investigated~\cite{mannos1974effects}. Recently, there has been a demonstrated success for developing the FR quality measures, including the Peak Signal-to-Noise Ratio (PSNR), Structural Similarity Index (SSIM) \cite{wang2004image}, Multiscale SSIM (MS-SSIM) \cite{wang2003multiscale}, Visual Salience-Induced Index (VSI) \cite{zhang2014vsi},  Median Absolute Deviation (MAD) \cite{larson2010most} and Visual Information Fidelity (VIF) \cite{sheikh2006image}. Unfortunately, in the vast majority of practical applications, the reference images are usually absent or difficult to obtain, leading to the exponential increase in the demand for no-reference (NR) IQA methods. Compared with FR-IQA, NR-IQA is a more challenging task due to the lack of pristine reference information.
%-Shiqi xxx add some FR popular models, and cite the reference I sent to you
%and usually cannot achieve the comparable performance as FR-IQA models. 

%--Shiqi herein, we need to use three or four sentences to summarize the NR-methods, including NSS, deep learning... 
{In the literature, numerous NR-IQA methods have been proposed based on the hypothesis that natural scenes possess certain statistical properties. Thus, the quality can be assessed by measuring the deviation of the  statistics between distorted and pristine images \cite{moorthy2011blind,saad2012blind, hou2014blind}.  With the development of deep learning technologies, the image quality can be inferred by learning from the labeled image data \cite{kang2014convolutional,kim2016fully,bosse2017deep,bianco2018use,gu2019blind,fu2016blind,kim2018multiple}. However, such data driven based methods highly rely on the large-scale training samples.}
Recently, the free-energy based brain theory \cite{friston2006free, friston2010free, gu2014using, zhai2011psychovisual} provides a novel solution for NR-IQA from the Bayesian view. \bl{In particular, the free energy theory reveals that human visual system (HVS)  may attempt to infer the reference signals to reduce the uncertainty of perceived visual signals by an internal inference model.}
%In particular, the free energy theory reveals that human visual system (HVS) always attempts to reduce the uncertainty and explains the scene of perceived visual stimulus by an internal generative model. 
%This theory can be intuitively verified by the example of assessing the quality of a noisy image. 
Rooted in the widely accepted view that the intrinsic, perceptually-meaningful and learnable features could govern the image quality, \bl{in this work, we focus on the feature level reference information estimation for IQA.}  This method avoids the modeling of the image signal space of which the understanding is still quite limited.  
%As shown in Fig.~\ref{pairmos}, even the same level noise is introduced in the two reference images, the quality degradation of the distorted image in the first row is much severe, while the  quality of the distorted image in the second row is nearly the same with its reference counterpart. This phenomenon reveals that  the effort we spend in restoring the reference image exactly may not necessary for some scenes. 
%In other words, as our ultimate goal is the quality prediction, what we really need to restore is the quality related reference features while the quality unrelated features which may be useful for reference image restoring will not contribute to the final quality regression. Therefore,
%Shiqi change the name to FPR...
Herein, we propose to learn a new NR-IQA measure named \textbf{FPR}, by inferring the quality through \textbf{F}eature-level \textbf{P}seudo-\textbf{R}eference  information. The underlying design philosophy of our method is learning the quality-specific PR feature instead of the restoration-specific PR feature. Along this vein, we can get rid of the design of a specific network for PR image generation, which is still a very challenging task. 
\bl{Besides, a gated recurrent units (GRU) \cite{chung2014empirical} aggregation strategy is proposed to aggregate the quality of each patch in the test image. The GRU network processes the image patches in order, and the long-term memory enjoyed by GRU can construct the long-dependency between each patch.}
%Moreover, comparing with the PR information constructed in feature-level, learning the PR in feature-level can also highly reduce the network scale and save the inference time in the testing phase. 
%The PR feature is learned  in a mutual manner. 
%More specifically, instead of learning the PR feature from a pre-trained FR-IQA network, both the NR-network and FR-IQA network are trainable. As such, more learnable reference feature can be generated by the FR-IQA network. 
To verify the performance of our method, we conduct both intra-database and cross-database experiments on four databases, including TID2013 \cite{ponomarenko2015image}, LIVE~\cite{sheikh2006statistical}, CSIQ \cite{larson2010most} and KADID-10k \cite{2019KADID}. Experimental results have demonstrated the superior performance of our method over existing state-of-the-art models. The main contributions of this paper are summarized as follows,
\begin{itemize}
\item \bl{We introduce the free-energy theory for the NR-IQA task  based on  deep learning. To capture the image distortion, the pristine feature is estimated by the supervision of an FR-IQA model, mimicking the internal inference mechanism in HVS.}

%We propose a novel NR-IQA framework based on the PR information constructed at the feature-level. This scheme aims to infer the pristine features that enjoy the advantages of quality-aware, learnable and discriminative. 

\item We learn the PR feature by a mutual learning strategy, leveraging the reference information. To improve the discrimination capability between the estimated PR feature and the distortion feature, a triplet loss is further adopted.

\item  We develop the aggregation strategy for the predicted scores of different patches in an image. The strategy benefits from the GRU, and generates the attention maps of the testing images for quality aggregation.

\end{itemize}

%The remainder of this paper is organized as follows. In Sec. II, we provide the related works of NR-IQA and mutual learning. Then we elaborate our proposed method in detail in Sec. III. The comparative studies of our FPR with state-of-the-art NR-IQA metrics are conducted in in Sec. IV.  Finally, some concluding remarks and future works are given in Sec. V.

%------------------------------------------------------------------------------------------
\section{Related Works}
Due to the lack of reference information, the existing NR-IQA measures can be classified into two categories: {quality-aware feature extraction based NR-IQA and discrepancy estimation based NR-IQA. In the first category, \bl{the quality-aware features are extracted based on a natural scene statistics (NSS) model or a data-driven model, and the quality}
%the quality-aware features are extracted based on a NSS model or a data-driven model, and the quality 
is finally predicted by a regression module.  In the second category, the PR images are first constructed, then the discrepancy between the input images and its PR images is measured. The philosophy is that the larger the discrepancy, the worse quality the image possesses.}  Herein, we provide an overview of the two categories of NR-IQA models as well as the mutual learning methods.

\subsection{Quality-Aware Feature Extraction based NR-IQA} 
{Typically, conventional NR-IQA methods extract the  quality-aware features based on the natural scene statistics (NSS) and predict the image quality by evaluating the destruction of naturalness. In \cite{mittal2012no,ye2013real}, based on the Mean-Subtracted Contrast-Normalized (MSCN) coefficients, the NSS are modeled with a generalized Gaussian distribution and the quality can be estimated by the distribution discrepancy.}  The NSS features have also been exploited in the wavelet domain,  \cite{moorthy2010two,hou2014blind,tang2011learning,wang2006quality}. In \cite{tang2011learning}, to discriminate degraded and undegraded images, the complex pyramid wavelet transform is performed and the magnitudes and phases of the wavelet coefficients are characterized as the NSS descriptor. {Analogously, in \cite{saad2012blind}, the discrete cosine transform (DCT) is introduced for NSS model construction, leading to a Bayesian inference-based NR-IQA model.} Considering the structural information is highly quality-relevant, the joint statistics of gradient magnitude and Laplacian of Gaussian response are utilized in  \cite{xue2014blind} to model the statistical naturalness. In \cite{sun21.mm}, the hybrid features consisting of texture components, structural information and intra-predicted modes are extracted and unified for adaptive bitrate estimation. Recently, there has been a surging interest in deep-feature extraction for NR-IQA. In \cite{kang2014convolutional}, a shallow ConvNet is first utilized for patch-based NR-IQA learning. This work is extended by DeepBIQ \cite{bianco2018use}, where a pre-trained convolutional neural network (CNN) is fine-tuned for the generic image description. Instead of learning only from the quality score, the multi-task CNN was proposed in \cite{kang2015simultaneous}, in which both the quality estimation and distortion identification are learned simultaneously for a better quality degradation measure. However, although those deep-learning based methods have achieved high-performance improvement, insufficient training data usually create the over-fitting problem. To alleviate this issue, extra synthetic databases  \textit{e.g.} \cite{zhang2018blind,ma2017end,chen2021no} have been proposed for more generalized model learning. The training data can also be enriched by ranking learning.  In \cite{liu2017rankiqa, niu2019siamese,ying2020quality,chen2021no}, the quality scores of an image pair are regressed and ranked, leading to the more quality-sensitive feature extraction. \bl{In~\cite{li2020norm,li2021unified}, to accelerate the model convergence,  “Norm-in-Norm” loss was proposed. The embedded normalization has been proved to be able to improve the smoothness of the loss landscape. It is generally acknowledged that combining different databases for IQA can highly enrich the training samples. However, the annotation shift usually causes unreliable data fusion. In~\cite{zhang2021uncertainty},  Zhang \textit{et al.} adopted rank-based learning to account for this problem, and the MOS uncertainty is also explored during optimization. The distortion type identification and quality regression are learned successively in ~\cite{yan2021precise}, aiming to capture more accurate distortion.}

\begin{figure*}[t]
\begin{minipage}[b]{1.0\linewidth}
  \centerline{\includegraphics[width=1\linewidth]{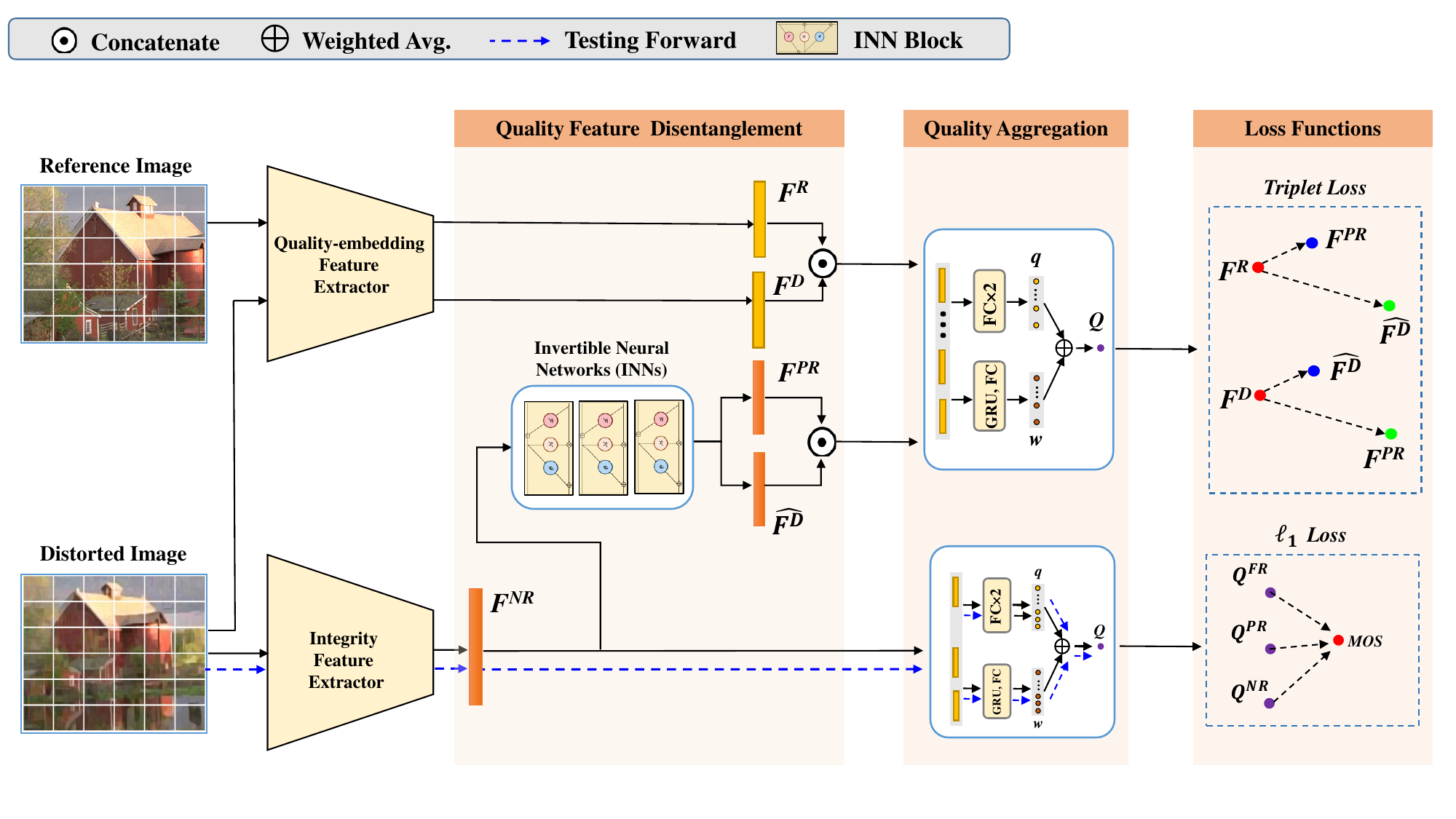}}
\end{minipage}
\caption{{Illustration of the framework of our proposed method. Two feature extractors are utilized in the training phase: the quality-embedding feature extractor and the integrity feature extractor. For the quality-embedding feature extractor, we extract the quality-embedding features  $F^{D}$ and  $F^{R}$ from the distortion image and its reference image, respectively.  For the integrity feature extractor, we aim to extract the  feature $F^{NR}$ from a single distorted image that contains the fusion information of a PR feature  $F^{PR}$ and a distortion feature $\hat{F^{D}}$ by the guidance of $ F^{R}$ and $F^{D}$. Then the quality of the distorted image can be regressed by the $F^{NR}$.  Finally, we  propose a \bl{GRU-based} quality aggregation module for patch-wise quality score aggregation. In the testing phase, only the testing image (without  reference) is available for quality prediction based on the proposed NR-IQA model. }}
\label{fig:frame}
\end{figure*}
\subsection{Discrepancy Estimation based NR-IQA}
The NR-IQA problem can be feasibly converted to the FR-IQA problem when the reference image can be inferred through generative models.
In \cite{lin2018hallucinated}, the PR image is generated by a quality-aware generative network, then the discrepancy between the distorted image and PR image is measured for quality regression. In contrary to constructing the PR image with perfect quality, the reference information provided by the PR image that suffered from the severest distortion was explored in \cite{min2017blind}, then an NR-IQA metric was developed by measuring the  similarity between the structures of distorted and the PR images. In \cite{jiang2020no}, both a pristine reference image (generated via a restoration model) and a severely distorted image (generated via a degradation model) are utilized for quality prediction. Analogously, by comparing the distorted images and their two bidirectional PRs, the bilateral distance (error) maps are extracted in \cite{hu2020tspr}.

\subsection{Mutual Learning} 
The assumption of mutual learning is highly relevant with the dual learning and collaborative learning, as their assumptions all lie in  the encouragement of the models to teach each other during training. For example,  the dual learning was adopted in \cite{he2016dual}, where two cross-lingual translation models are forced to learn from each other through the reinforcement learning. Comparing with dual learning, the same task is learned in the collaborative learning. In \cite{batra2017cooperative}, multiple student models are expected to learn the same classification task while their inputs are sampled from different domains. Different from the dual learning and collaborative learning, both the tasks and inputs of the models in mutual learning are identical. For example, the deep mutual learning was utilized in \cite{zhang2018deep}, where two student models are forced to learn the  classification collaboratively by a Kullback Leibler (KL) loss. This work was further extended in \cite{lai2019adversarial}, with the KL loss replaced by a generative adversarial network. {In our method, the mutual learning strategy was adopted to improve the learnability of the PR feature and we further impose the triplet constraint to the output features, significantly enhancing their discriminability.}

%------------------------------------------------------------------------------------------

\section{The Proposed Scheme}
We aim to learn an NR-IQA measure by hallucinating the PR features. In the training stage, given the pristine reference, we attempt to build a FR-IQA model with the distorted and corresponding pristine reference images. The PR feature is subsequently learned in a mutual way. Finally, the \bl{GRU-based} quality aggregation is performed to obtain the final quality score. 

%As shown in Fig.~\ref{fig:frame}, our method consists of three steps, including: 1) Mutual learning for PR feature generation. 2) GRU based quality aggregation. 3) Objective functions defining for model training. The details of above three steps will be elaborated in the following subsections. 

%------------------------------------------------------------------------------------------
\subsection{PR Feature Learning}

%To construct the PR feature, a trivial way is to pre-train a FR-IQA model, then the PR feature can be learned by minimize the distance between the feature generated from the distorted images and the feature generated from the reference feature. However, we argue this method may not effective enough. 
%Shiqi: what is the NR-Network? is it well defined?
%Shiqi: explain the training stage: what information you have, the general philosophy 

{As shown in Fig.~\ref{fig:frame}, in the training phase, we learn to hallucinate the PR feature $F^{PR}$ from a single distorted image by the guidance of the pristine reference feature $F^{R}$. \bl{In particular, the $F^{R}$ is constructed from an FR model based on the quality-embedding feature extractor.}
%In particular, the $F^{R}$ is generated from a FR model based on the quality-embedding feature extractor 
and the $F^{PR}$ is decomposed from the feature $F^{NR}$ which is regarded as a fusion feature that contains the entire information of the PR feature $F^{PR}$  and the distortion feature $\hat{F^{D}}$.}

In general, there are two properties a desired PR feature for quality prediction should possess. First, the PR feature should be learnable. \bl{Constructing the pristine image from the distorted one is usually a challenging task due to the corruption of content caused by different distortion types. For example, the texture regions could be difficult to be recovered when it is corrupted by blur distortion. Such challenges also bring the difficulties to learn pristine (PR) features.  To account for this, self-supervised based methods can be adopted, such as the generative adversarial network (GAN) based models ~\cite{kim2020global,chen2018deep} or natural images prior based methods ~\cite{noroozi2016unsupervised,zhang2021self}. As indicated in ~\cite{jing2020self}, the self-supervised learning methods greatly depend on the difficulty level of the task and the amount of training data. In this paper, we learn the PR feature in a supervised manner based on the reference as it is possible to obtain abundant pristine images. In particular,  the features extracted from the reference images are used as guidance to provide a clear picture for the PR feature learning.}
%Generally speaking, it is an ill-posed problem for learning the PR information from the distorted image. 
Though in the training process the reference image could be taken advantage of, the PR feature may not be able to be feasibly learned by only forcing the inferred PR feature to be close to a pre-defined reference feature. As such, \bl{the learning capability of the NR-IQA network should be carefully considered during the reference feature estimation.} In other words, when we learn the PR feature, the  reference feature extraction should be learned mutually.
%the learning capability of the NR-IQA network should be carefully considered during the reference feature generation. 
%Otherwise, the uneasy learnable reference feature will cause an inaccurate PR feature estimation, leading to a poor performance on final quality prediction. 
Second, the PR feature should be discriminative enough when comparing with the distortion feature. Enhancing the discriminability could improve the quality sensitivity of the PR feature and subsequently promote the prediction performance.

%In our method, the PR feature learning scheme is designed by taking these two properties into full consideration. 

%In particular, we first crop the input reference image and distorted image into different patches with the size set by $64 \times 64$. This is a common operation for IQA as the sizes of input images can be vary. Further more, the patch cropping will also enlarge the number of training samples, alleviating the over-fitting problems.  --Shiqi need to be added somewhere
\begin{figure}
\begin{center}
\includegraphics[width=0.7\textwidth]{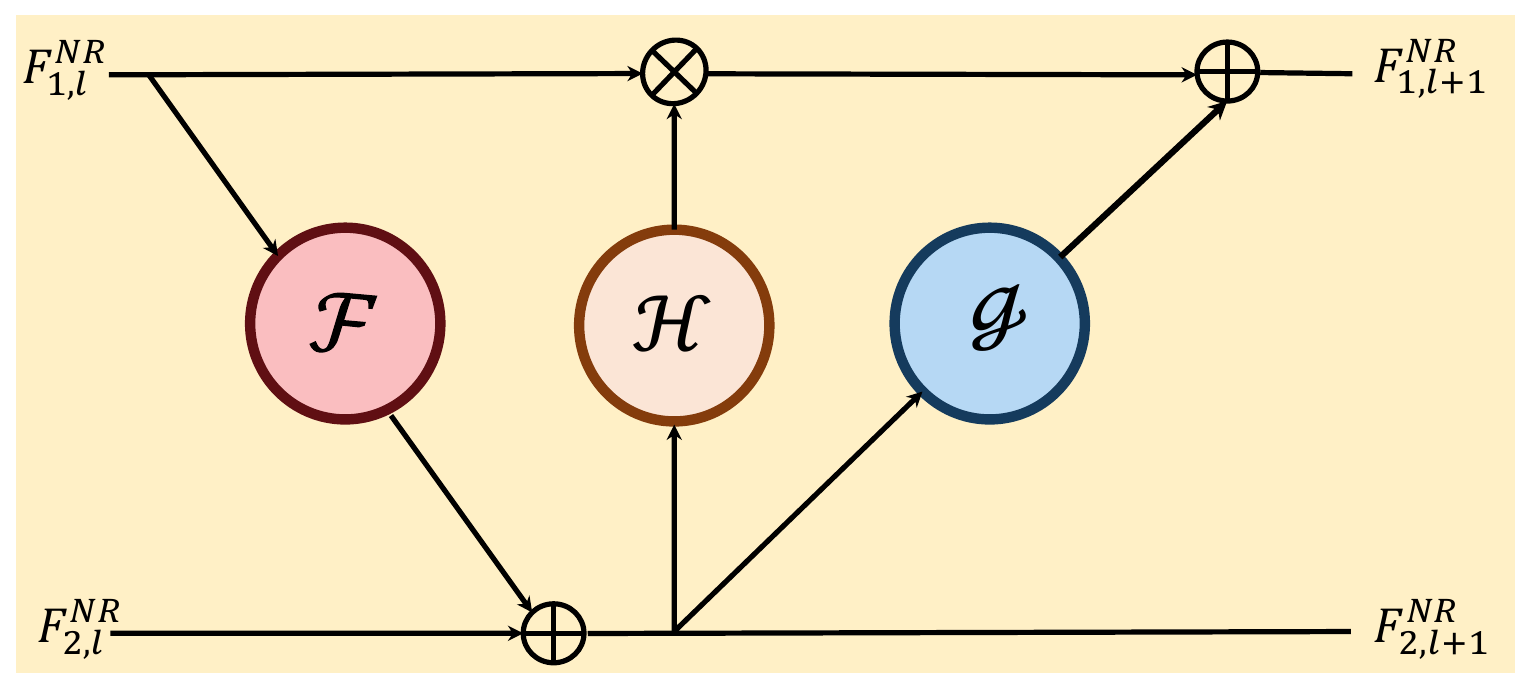}
\caption{\bl{The structure of invertible neural network.}}
\label{fig:innNet}
\end{center}
\end{figure}

The proposed method is conceptually appealing in the sense of learnability and discriminability. Regarding the learnability, a mutual learning strategy is adopted. 
%, attempting to learn rather than learn the PR feature from a pre-trained NR-model. 
%To generate the PR feature, we input 
As shown in Fig.~\ref{fig:frame}, in the training process, the paired images including the distorted image and its corresponding reference are fed into the quality-embedding feature extractor,  %two feature extractors named FR feature extractor and NR feature extractor are adopted subsequently. 
generating the reference feature $F^{R}$ and distortion feature $F^{D}$. 
\bl{The integrity feature extractor, which accepts the distorted image only, is encouraged to generate the feature $F^{NR}$ with high quality-awareness. To this end, we force the $F^{NR}$ to be derived from both the information of of the pseudo reference feature  $F^{PR}$ and  the distortion feature  $\hat{F^{D}}$. As shown in Fig. \ref{fig:frame}, we adopt the invertible neural networks (INNs) \cite{dinh2014nice,dinh2016density} to disentangle the $F^{NR}$ into a pseudo reference feature  $F^{PR}$ and a distortion feature  $\hat{F^{D}}$, without losing any information. Instead of using the concatenation of  $F^{PR}$ and  $\hat{F^{D}}$ for quality regression, the $F^{NR}$ utilized enjoys higher quality awareness, generalization capability, and less inference time. In Fig. \ref{fig:innNet}, we plot the structure of an INN block, which consists of transmission functions including $\mathcal{F}$, $\mathcal{H}$, and $\mathcal{G}$. For the $l$-th block, the  $F^{NR}_{l}$ is split into $F^{NR}_{1,l}$ and  $F^{NR}_{2,l}$ along the channel axis, and they undergo the invertible  transformations ~\cite{dinh2014nice,dinh2016density} as follows,
\begin{equation}\label{inn1}
\begin{aligned}
&F^{NR}_{1,l+1}=F^{NR}_{1,l}+\mathcal{F}\left(F^{NR}_{2,l}\right), \\
&F^{NR}_{2,l+1}= F^{NR}_{2,l} \odot \exp \left(\mathcal{H}\left(F^{NR}_{1,l+1}\right)\right)+\mathcal{G}\left(F^{NR}_{1,l+1}\right). \\
\end{aligned}
\end{equation}
The inverse transformation is computed as follows,
\begin{equation}\label{inn2}
\begin{aligned}
&F^{NR}_{2,l}= \left(F^{NR}_{2,l+1}-\mathcal{G}\left(F^{NR}_{1,l+1}\right)\right)\odot \exp \left(-\mathcal{H}\left(F^{NR}_{1,l+1}\right)\right) \\
&F^{NR}_{1,l}=F^{NR}_{1,l+1}-\mathcal{F}\left(F^{NR}_{2,l}\right).
\end{aligned}
\end{equation}
The outputs $F^{NR}_{1,l+1}$ and $F^{NR}_{1,l+1}$  are fed to the following INN blocks, and three blocks are finally utilized. We denote the output two features at the third block as $F^{PR}$ and $\hat{F^{D}}$ which are constrained by the triplet loss $\mathcal{L}_{trip}$ for the  discriminative reference feature learning.}

%In this step, two learning strategies we can adopt. The first is pre-training the FR feature extractor to generate the $F^{R}$ and $F^{D}$, then we fix the parameters of  the FR feature extractor and reduce the distance between reference features ($F^{R}$, $F^{PR}$ ) and distortion  features ($F^{D}$ and $\hat{F^{D}}$) by a distance measure introduced. However, we argue this is not an optimal way, due to the reference feature $F^{R}$ is constructed directly from the reference information, it will be extremely hard to restore the $F^{R}$ from the distorted images, especially when the distortion level is high. The key issue is the  learnability of the NR feature extractor is not considered during the pre-traning phase of FR feature extractor. To count for this, 
The mutual learning strategy enables the integrity feature extractor and quality-embedding feature extractor to be learned simultaneously with the feature distance constraint. Thus, more learnable reference feature can be generated by the FR model. The connection among the features $F^{NR}$, $F^{PR}$ and $\hat{F^{D}}$ are constructed by an invertible layer, consisting of three invertible neural networks (INNs) \cite{ardizzone2018analyzing}.  
%Herein, comparing with the ``post fusion'' strategy like concatenating the $F^{PR}$ and $\hat{F^{D}}$ directly, we find the ``pre-fusion'' strategy that learn the fusion mechanism implicitly and directly from the input distorted images can highly improve the performance of our model which has been verified by our experiments and we will elaborate it in Sec.IV. 
Through the INNs, the integrity feature  $F^{NR}$ can be disentangled in to a pseudo reference feature  $F^{PR}$ and  a distortion feature  $\hat{F^{D}}$, without losing any information due to the invertibility of INNs. 
%Herein, our goal now is to reduce the discrepancy between $F^{R}$ and $F^{PR}$ and  the discrepancy between  $F^{D}$ and $\hat{F^{D}}$. 

To equip the discriminative capability, a triplet loss is further utilized \cite{schroff2015facenet} as the distance measure between the reference features ($F^{R}$, $F^{PR}$ ) and the corresponding distortion features ($F^{D}$ and $\hat{F^{D}}$), which is expressed as follows,
\begin{equation}\label{trip}
\begin{aligned}
\mathcal{L}_{trip}= \sum_{i=1}^{N}\left[\left\|F^{R}_i-F^{PR}_i\right\|_{2}^{2}-\left\|F^{R}_i-\hat{F^{D}}_i\right\|_{2}^{2}+\delta\right]_{+} \\
+\sum_{i=1}^{N}\left[\left\|F^{D}_i-\hat{F^{D}}_i\right\|_{2}^{2}-\left\|F^{D}_i-F^{PR}_i\right\|_{2}^{2}+\delta\right]_{+},
\end{aligned}
\end{equation}
where $i$ is input patch index in a batch, $N$ is the batch size and $\delta $ is a margin that is enforced between positive and
negative pairs. {With this loss, on the one hand, the distance between the reference feature and PR feature can be reduced. On the other hand, the discrepancies between the reference/PR feature and two distortion features can be enlarged.

As illustrated in Fig.~\ref{fig:frame}, {to maintain the relationship of $F^{PR}$ and $\hat{F^{D}}$ to be consistent with  $F^{R}$ and $F^{D}$}, we  concatenate $F^{R}$ with $F^{D}$ (denoted as $Concat(F^{R}, F^{D})$) and $F^{PR}$ with $\hat{F^{D}}$ (denoted as $Concat(F^{PR}, \hat{F^{D}})$) for quality prediction through a shared quality aggregation module.}

%------------------------------------------------------------------------------------------
\iffalse
\begin{figure}[t]
\begin{minipage}[b]{1.0\linewidth}
  \centerline{\includegraphics[width=1\linewidth]{Figs/agg.pdf}}
\end{minipage}
\caption{{Different quality aggregation strategies. (a) Equal-weight average. (b) Saliency map guided aggregation.  (c) Patch-wise weight generation based aggregation. (d) Ours \bl{GRU-based} weighted average. }}
\label{fig:agg}
\end{figure}
\fi

\subsection{GRU-Based Quality Aggregation}
%Due to the input of our network is patch-based, when the sizes of the images are vary, the number of patches will be different. 
To aggregate the predicted quality score of each patch in an image, the aggregation module should be invariant of the patch numbers. 
%In Fig~\ref{fig:agg}, four different aggregation strategies are presented. The first is pooling-by-average strategy which is shown in the Fig.~\ref{fig:agg} (a). This is a trivial way to fuse the score of each patch, as human usually perceive different regions in an image with the different amount of attention. To account for this, the attention map estimation is adopted to aggregate the local qualities by different weights \cite{zhang2015application,zhang2014vsi}. As shown in Fig.~\ref{fig:agg} (b), a saliency map estimation module is introduced into image quality aggregation. However, the pre-trained saliency models incorporated are mostly generated from the reference image or a combination of reference image and distorted image, limiting its application on the NR-IQA task. To learn the attention map and quality prediction explicitly in a joint optimization manner, Bosse \textit{et al.} \cite{bosse2017deep} propose to generate the patch-wise weight by 
%integrating another branch in parallel to the patch-wise quality regression branch which is shown in Fig.~\ref{fig:agg} (c). However, the weight of each patch is estimated only from the feature generated by the corresponding one patch, the perceptual relevance between different local patches are not considered. 
In this paper, we propose a \bl{GRU-based} quality score aggregation module as shown in Fig.~\ref{fig:frame}. More specifically, regarding the concatenated features $Concat(F^{R}, F^{D})$ and $Concat(F^{PR}, \hat{F^{D}})$, two sub-branches are adopted for quality prediction. The first branch is a fully-connected (FC) layer that is responsible for patch-wise quality prediction with the patch-wise concatenated feature as input. Another sub-branch consists of one GRU layer and one FC layer. Different from  \cite{bosse2017deep}, the inputs of the GRU layer are the features of all the patches in an image. With GRU, the long-term dependencies between different patches can be modeled and synthesized, then we normalize the output weights from the last FC layer for final attention map generation,
\begin{equation}\label{nor}
w_{i}=\frac{\alpha_{i}}{\sum_{j=1}^{N_{p}} \alpha_{j}},
\end{equation}
where $i$ is the patch index in an image, and $N_p$ is the number of patches.
% and $p_i$ is the attention weight of $i-{th}$ patch.  
$\alpha_i$ and $w_i$ are the predicted and normalized attention weights of $i-$th patch, respectively. Finally, the global image quality $Q$ can be estimated as
\begin{equation}\label{iq}
Q =\sum_{i=1}^{N_{p}} w_{i} q_{i}=\frac{\sum_{i=1}^{N_{p}} \alpha_{i} q_{i}}{\sum_{i=1}^{N_{p}} \alpha_{i}},
\end{equation}
where $q_i$ is the predicted quality of $i-{th}$ patch.  As shown in  Fig.~\ref{fig:frame},  we also adopt the same  strategy (network) for the quality aggregation of the patch-wise integrity feature $F^{NR}$. Due to the distinct representations of the fused feature and concatenated features, the parameters of the two aggregation modules are not shared during the model training. 

%------------------------------------------------------------------------------------------
\subsection{Objective Function}
In summary, the objective function in our proposed method includes two triplet losses and three quality regression losses. In particular, for quality regression, compared with mean squared error (MSE), optimization with mean absolute error (MAE) is less sensitive to the outliers, leading to a more stable training processing. Consequently, the objective function is given by,
\bl{
\begin{equation}\label{all}
\begin{aligned}
\mathcal{L} 
&=  \mathcal{L}_{mae} + \lambda \mathcal{L}_{trip} \\
&= \sum_{j=1}^{N_I}\left[ \left|\hat{Q_j}-Q^{FR}_{j}\right|+ \left|\hat{Q_j}-Q^{PR}_{j}\right|+ \left|\hat{Q_j}-Q^{NR}_{j}\right|\right]\\
& \quad  +\lambda \left(\sum_{i=1}^{N_I*N_P}\left[\left\|F^{R}_i-F^{PR}_i\right\|_{2}^{2}-\left\|F^{R}_i-\hat{F^{D}_i}\right\|_{2}^{2}+\delta\right]_{+}\right) \\
&\quad + \lambda \left(\sum_{i=1}^{N_I*N_P}\left[\left\|F^{D}_i-\hat{F^{D}_i}\right\|_{2}^{2}-\left\|F^{D}_i-F^{PR}_i\right\|_{2}^{2}+\delta\right]_{+} \right),
\end{aligned}
\end{equation}}
where $i$ and $j$ are the patch index and image index, respectively. $N_I$ is the number of images in a batch and $\hat{Q_j}$ is the Mean Opinion Score (MOS) provided by the training set. $Q^{FR}$, $Q^{PR}$, $Q^{NR}$ are the quality scores predicted from the features $Concat(F^R, F^D)$, $Concat(F^{PR}, \hat{F^{D}})$ and $F^{NR}$, respectively.   \bl{In addition, we adopt $\lambda$ as the weight of the regularization term ($\mathcal{L}_{trip}$)  for quality regression.} It is also worth noting that the extractions of $F^R, F^D,F^{PR}, \hat{F^{D}}$ are not necessary in the testing phase, and we only adopt the $Q^{NR}$ for the final quality prediction, thus the {computational complexity} in testing phase can be highly reduced comparing with the network used in the training phase.
%-Shiqi what is network complexity?

%In principle, we can crop the image into patches with different overlapping pixels. However, we find there is no significant performance drop when the patches are cropped without overlapping both in training and testing phases. 
%------------------------------------------------------------------------------------------
\begin{table*}
  \centering
  \caption{Descriptions of the Four IQA databases.}
\setlength{\tabcolsep}{1.8mm}{
\begin{tabular}{c|m{4.0em}<{\centering}|c|m{35.365em}<{\centering}}
\hline
Database & \multicolumn{1}{m{4.0em}<{\centering}|}{\# of Ref. Images} & \multicolumn{1}{m{3.5em}<{\centering}|}{\# of Images} & Distortion Types \bigstrut\\
\hline
TID2013~\cite{ponomarenko2015image} & 25    & 3,000  & Additive Gaussian noise; Additive noise in color components; Spatially correlated noise; Masked noise; High frequency noise; Impulse noise; Quantization noise; Gaussian blur; Image denoising; JPEG compression; JPEG2000 compression; JPEG transmission errors; JPEG2000 transmission errors; Non eccentricity pattern noise; Local block-wise distortions of different intensity; Mean shift (intensity shift); Contrast change; Change of color saturation; Multiplicative Gaussian noise; Comfort noise; Lossy compression of noisy images; Image color quantization with dither; Chromatic aberrations; Sparse sampling and reconstruction \bigstrut\\
\hline
LIVE~\cite{sheikh2003image}  & 29    & 982   & JPEG and JPEG2000 compression; Additive white; Gaussian noise; Gaussian blur; Rayleigh fast-fading channel distortion   \bigstrut\\
\hline
CSIQ~\cite{larson2010most}  & 30    & 866   & JPEG compression; JP2K compression; Gaussian blur; Gaussian white noise; Gaussian pink noise and contrast change \bigstrut\\
\hline
KADID-10k~\cite{2019KADID}  & 81    & 10,125 &   Blurs (Gaussian, lens,motion); Color related (diffusion, shifting, quantization, over-saturation and desaturation); Compression (JPEG2000, JPEG); Noise related (white, white with color, impulse, multiplicative white noise + denoise); Brightness changes (brighten, darken, shifting the mean); Spatial distortions (jitter, non-eccentricity patch, pixelate, quantization, color blocking); Sharpness and contrast. \bigstrut\\
\hline
\end{tabular}%

}
 \label{tab:list}%
\end{table*}%

\section{Experimental Results}

\subsubsection{IQA Databases}
Since our model is trained in a paired manner, the reference image should be available during the training phase. As such, to validate the proposed method, we evaluate our model on four synthetic natural databases  including: TID2013~\cite{ponomarenko2015image}, LIVE~\cite{sheikh2003image}, CSIQ~\cite{larson2010most} and KADID-10k~\cite{2019KADID}. More details are provided in Table~\ref{tab:list}.

\textbf{TID2013.} The TID2013 database consists of 3,000 images obtained from 25 pristine images for reference. The pristine images are corrupted by 24 distortion types and each distortion type corresponds to 5 levels. The image quality is finally rated by double stimulus procedure and the MOS values are obtained in the range [0, 9], where larger MOS indicates better visual quality. 
 
\textbf{LIVE.} The LIVE IQA database includes 982 distorted natural images and 29 reference images. Five different distortion types are included: JPEG and JPEG2000 compression, additive white Gaussian noise (WN), Gaussian blur (BLUR), and Rayleigh fast-fading channel distortion (FF).  Different from the construction of TID2013, a single-stimulus rating procedure is adopted for quality rating, producing a range of different mean opinion scores (DMOS) from 0 to 100 and a lower DMOS value represents  better image quality.

\textbf{CSIQ.} The CISQ database contains 30 reference images and 866 distorted images. This database involves six distortion types: JPEG compression, JP2K compression, Gaussian blur, Gaussian white noise, Gaussian pink noise and contrast change. The images are  rated by 35 different observers and the DMOS results are normalized into the range [0, 1].

\textbf{KADID-10k.} In this database, 81 pristine images are included and each pristine image is degraded by 25 distortion types in 5 levels.  All the images are resized into the same resolution (512×384). For each distorted image, 30 reliable degradation category ratings have been obtained by crowdsourcing. 

%In addition, the distortion types included in above four databases are presented in Table~\ref{tab:list} for better comparison.

\begin{figure*}[t]
\begin{minipage}[b]{1.0\linewidth}
  \centerline{\includegraphics[width=1\linewidth]{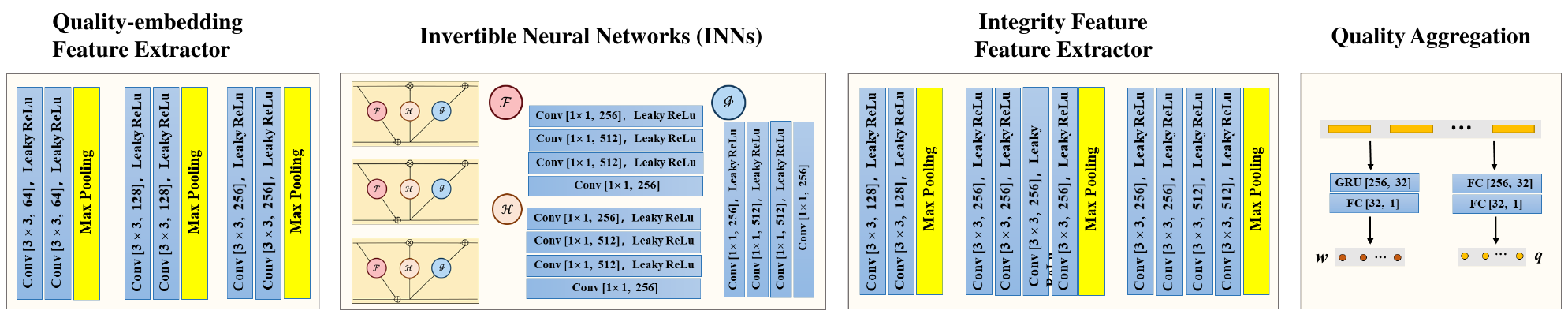}}
\end{minipage}
\caption{{Illustration of the network architectures for the quality-embedding feature extractor, INNs, integrity feature extractor and quality aggregation module. }}
\label{fig:netw}
\end{figure*}

\subsubsection{Implementation Details}
We implement  our  model  by  PyTorch~\cite{paszke2019pytorch}. In Fig.~\ref{fig:netw}, we show the layer-wise network design of our proposed method. We crop the image patches without overlapping and the size is set by $64\times 64$. The number of image pairs in a batch is set by 32. 
%In principle, we can crop the image into patches with different overlapping pixels. However, we find there is no significant performance drop when the patches are cropped without overlapping in both training and testing phases. 
We adopt Adam optimizer~\cite{kingma2014adam} for optimization. The learning rate  is fixed to 1e-4 with a weight decay set as 1e-4. The weighting parameters $\lambda, \delta$ in Eqn.~\eqref{all} are set as 20.0 and 0.5, respectively. We duplicate the  samples by 16 times in a batch to augment the data. The maximum epoch is set by 1,000. 

It should be mentioned that all the experimental pre-settings are fixed both in intra-database and cross-database training. For the intra-database evaluation, we randomly split the  dataset into the training set, validation set and testing set by reference image to guarantee there is no content overlap among the three sets. In particular, 60\%, 20\%, 20\% images are used for training, validation and testing, respectively. We discard the 25th reference image and the distorted versions in TID2013, as they are not natural image.  The experimental results on intra-database are reported based on 10 random splits. To make errors and gradients comparable for different databases, we linearly map the MOS/DMOS ranges of the other three databases (TID2013, CSIQ, KADID-10k) to the DMOS range [0, 100] which is the same as LIVE database. \bl{Three evaluation metrics are reported for each experimental setting, including Spearman rank correlation coefficient (SRCC), Pearson linear correlation coefﬁcient (PLCC) and perceptually weighted rank correlation (PWRC) \cite{wu2018perceptually}. The  PLCC  evaluates the prediction accuracy and the SRCC indicates the prediction monotonicity. Regarding the PWRC, the perceptual importance variation and subjective uncertainty are considered, which is confirmed to be reliable in recommending the perceptually preferred IQA model.}

\begin{table}
  \centering
  \caption{\bl{Performance evaluation on the LIVE, CSIQA and TID2013 databases. The top two results are highlighted in boldface.}}
\resizebox{250pt}{95pt}{
% Table generated by Excel2LaTeX from sheet 'Sheet2'
\begin{tabular}{l|cc|cc|cc}
\hline
\multirow{2}[4]{*}{Method} & \multicolumn{2}{c|}{LIVE} & \multicolumn{2}{c|}{CSIQ} & \multicolumn{2}{c}{TID2013} \bigstrut\\
\cline{2-7}      & SRCC  & PLCC  & SRCC  & PLCC  & SRCC  & PLCC \bigstrut\\
\hline
BRISQUE~\cite{mittal2012no} & 0.939 & 0.935 & 0.746 & 0.829 & 0.604 & 0.694 \\
M3~\cite{xue2014blind}    & 0.951 & 0.950  & 0.795 & 0.839 & 0.689 & 0.771 \\
FRIQUEE~\cite{ghadiyaram2017perceptual}  & 0.940  & 0.944 & 0.835 & 0.874 & 0.680  & 0.753 \\
CORNIA~\cite{ye2012unsupervised}  & 0.947 & 0.950  & 0.678 & 0.776 & 0.678 & 0.768 \\
HOSA~\cite{xu2016blind}  & 0.946 & 0.947 & 0.741 & 0.823 & 0.735 & 0.815 \\
Le-CNN~\cite{kang2014convolutional}  & 0.956 & 0.953 &   -    &   -    &   -    & - \\
BIECON~\cite{kim2016fully}  & 0.961 & 0.962 & 0.815 & 0.823 & 0.717 & 0.762 \\
DIQaM-NR~\cite{bosse2017deep}  & 0.960  & 0.972 &  -     &     -  & 0.835 & 0.855 \\
WaDIQaM-NR~\cite{bosse2017deep} & 0.954 & 0.963 & -      &  -     & 0.761 & 0.787 \\
ResNet-ft~\cite{kim2017deep}  & 0.950  & 0.954 & 0.876 & 0.905 & 0.712 & 0.756 \\
IW-CNN  ~\cite{kim2017deep} & 0.963 & 0.964 & 0.812 & 0.791 & 0.800   & 0.802 \\
DB-CNN ~\cite{kim2017deep} & 0.968 & 0.971 & \textbf{0.946} & \textbf{0.959} & 0.816 & 0.865 \\
CaHDC~\cite{wu2020end} & 0.965 & 0.964 & 0.903 & 0.914 &  0.862 &  0.878 \\
HyperIQA~\cite{wu2020end} & 0.962 & 0.966 & 0.923 & 0.942 &  0.729     & 0.775\\

Hallucinated-IQA ~\cite{lin2018hallucinated}  & \textbf{0.982} & \textbf{0.982} & 0.885 & 0.910 &  \textbf{0.879}     & \textbf{0.880} \\
%LinearityIQA ~\cite{li2020norm} & 0.963 & 0.967 & 0.896 & 0.908 &  0.786     & 0.834 \\
%UNIQUE~\cite{zhang2021uncertainty} & \textbf{0.969} & 0.968 & 0.902 & 0.927 &  -     & -\\
\hline
FPR (Ours)   &  \textbf{0.969}     &  \textbf{0.974}     &   \textbf{0.950}
    &  	\textbf{0.958}     &  \textbf{0.854}
     & 	\textbf{0.882} \\
FPR (FR) &   0.971 &    0.982   &   0.966 &   0.969    &    0.893 &  0.876\\
\hline
\end{tabular}%

}
  \label{tab:three}%
\end{table}%

% Table generated by Excel2LaTeX from sheet 'kadid'
\begin{table*}[htbp]
  \centering
  \caption{\bl{Performance evaluation on the KADID-10k database. The top two results are highlighted in boldface.}}
    \resizebox{512pt}{40pt}{
    \begin{tabular}{p{4.045em}|ccccccc}
    \toprule
    Method & {BIQI~\cite{moorthy2010two} } & {BLIINDS-II~\cite{saad2012blind}  } & {BRISQUE~\cite{mittal2012no} } & {CORNIA~\cite{ye2012unsupervised}} & {DIIVINE~\cite{moorthy2011blind} } & {HOSA~\cite{xu2016blind}} & {SSEQ~\cite{liu2014no}} \\
    \midrule
    SROCC & 0.431  & 0.527  & 0.519  & 0.541  & \multicolumn{1}{c}{0.489 } & 0.609  & 0.424  \\
    PLCC  & 0.460  & 0.559  & 0.554  & 0.580  & \multicolumn{1}{c}{0.532 } & 0.653  & 0.463  \\
    \midrule
    Method & {InceptionResNetV2~\cite{2019KADID}} & {WaDIQaM-NR~\cite{bosse2017deep}} & {DB-CNN ~\cite{kim2017deep} } &{HyperIQA~\cite{wu2020end}} & {LieartyIQA~\cite{li2020norm}} & {FPR (Ours)} & {FRP (FR)} \\
    \midrule
    SROCC & 0.731  & \textbf{0.845 } & 0.843  & 0.818  & 0.828  & \textbf{0.894 } & 0.916  \\
    PLCC  & 0.734  & \textbf{0.851 } & 0.845  & 0.818  & 0.818  & \textbf{0.898 } & 0.911  \\
    \bottomrule
    \end{tabular}%
    }
  \label{tab:kad}%
\end{table*}%

\begin{table}
  \centering
  \caption{\bl{PWRC comparison in four IQA databases. The top two results are highlighted in boldface.}}
 % Table generated by Excel2LaTeX from sheet 'Sheet9'
        \resizebox{250pt}{68pt}{
        \begin{tabular}{l|cccc}
        \midrule
        Method & CSIQ  & KADID-10k & LIVE  & TID2013 \\
        \midrule
        NIQE  ~\cite{mittal2012making}  & 19.056  & 9.482  & 5.379  & 1.449  \\
        BRISQUE ~\cite{mittal2012no}& 26.940  & 13.111  & 5.725  & 2.663  \\
        FRIQUEE ~\cite{ghadiyaram2017perceptual}& 25.515  & 22.454  & \textbf{5.860 } & 4.031  \\
        ILNIQE   ~\cite{zhang2015feature}  & 25.695  & 14.989  & 5.216  & 2.640  \\
        DIQaM-NR ~\cite{bosse2017deep} & 25.528  & 17.941  & 4.915  & 3.019  \\
        WaDIQaM-NR ~\cite{bosse2017deep}& 27.221  & \textbf{25.205 } & 5.770  & 4.547  \\
        DB-CNN  ~\cite{kim2017deep} & \textbf{29.312 } & 21.805  & 5.810  & \textbf{4.634 } \\
        HyperIQA ~\cite{wu2020end} & 26.388  & 22.214  & 5.471  & 4.386  \\
        LieartyIQA ~\cite{li2020norm} & 25.300  & 21.285  & 5.736  & 4.492  \\
        \midrule
        FPR (Ours) & \textbf{30.328 } & \textbf{25.492 } & \textbf{5.920 } & \textbf{5.515 } \\
        \bottomrule
        \end{tabular}%
        }
      \label{tab:pwrc}%
\end{table}%

\begin{table*}
  \centering
  \caption{\bl{SRCC comparison in three databases on four common distortion types. The top two results are highlighted in boldface.}}
 \resizebox{512pt}{60pt}{
% Table generated by Excel2LaTeX from sheet 'pr'
\begin{tabular}{c|c|cccccccccc|c}
\hline
\multirow{2}[4]{*}{Dataset} & \multirow{2}[4]{*}{Dist.Type} & \multicolumn{11}{c}{Method} \bigstrut\\
\cline{3-13}      &       & DIIVINE~\cite{moorthy2011blind} & BLINDS-II~\cite{saad2012blind} & BRISQUE~\cite{mittal2012no} & CORNIA~\cite{ye2012unsupervised} & HOSA~\cite{xu2016blind}  & WaDIQaM-NR~\cite{bosse2017deep} & DIQA~\cite{kim2018deep}   & BPRI (c)~\cite{min2017blind}  & BPRI (p)~\cite{min2017blind}  & TSPR~\cite{hu2020tspr}  & FPR (Ours) \bigstrut\\
\hline
\multirow{4}[2]{*}{LIVE} & WN    & \textbf{0.988} & 0.947 & 0.979 & 0.976 & 0.975 & 0.970 & \textbf{0.988}  & 0.984 & 0.985 & 0.972 & \textbf{0.987} \bigstrut[t]\\
      & GB    & 0.923 & 0.915 &  0.951 & 0.969 & 0.954 & 0.960 & 0.962  & 0.927 & 0.924 & \textbf{0.978} & \textbf{0.979}  \\
      & JPEG  & 0.921 & 0.950 & 0.965 & 0.955 & 0.954 & 0.964 & \textbf{0.976}  & \textbf{0.967} & \textbf{0.967} & 0.947 &  0.932\\
      & JP2K  & 0.922 & 0.930 & 0.914 & 0.943 & 0.954 & 0.949 & \textbf{0.961} & 0.908 & 0.907 & 0.950 &  \textbf{0.965}\bigstrut[b]\\
\hline
\multirow{4}[1]{*}{CSIQ} & WN    & 0.866 & 0.760 & 0.682 & 0.664 & 0.604 & 0.929 & 0.835 & 0.931 & \textbf{0.936} & 0.910 & \textbf{0.942} \bigstrut[t]\\
      & GB    & 0.872 & 0.877 & 0.808 & 0.836 & 0.841 & \textbf{0.958} & 0.870  & 0.904 & 0.900   & 0.908 &  \textbf{0.939}\\
      & JPEG  & 0.800 & 0.899 & 0.846 & 0.869 & 0.733 & 0.921 & 0.931  & 0.918 & 0.930  & \textbf{0.944} & \textbf{0.969} \\
      & JP2K  & 0.831 & 0.867 & 0.817 & 0.846 & 0.818 & 0.886 & \textbf{0.927} & 0.863 & 0.862 & 0.896 & \textbf{0.967} \\
\hline
\multirow{4}[1]{*}{TID 2013} & WN    & 0.855 & 0.647 & 0.858 & 0.817 & 0.817 & 0.843 & 0.915  & \textbf{0.918} & \textbf{0.918} & 0.876 & \textbf{0.931} \\
      & GB    & 0.834 & 0.837 & 0.814 & 0.840 & 0.870 & 0.861 & \textbf{0.912}  & 0.873 & 0.859 & 0.837 & \textbf{0.912} \\
      & JPEG  & 0.628 & 0.836 & 0.845 & 0.896 & \textbf{0.986} &  \textbf{0.931} & 0.875 & 0.907 & 0.910  & 0.913 & 0.906 \\
      & JP2K  & 0.853 & 0.888 & 0.893 & 0.901 & 0.902 &  \textbf{0.932} & 0.912 & 0.883 & 0.868 & \textbf{0.935} &  0.900\bigstrut[b]\\
\hline
\end{tabular}%

}
  \label{tab:FPR}%
\end{table*}%

\subsection{Quality Prediction on Intra-database}
\subsubsection{Overall Performance on Individual Database}
In this sub-section, we compare our method with other state-of-the-art NR-IQA methods, including BRISQUE~\cite{mittal2012no}, M3~\cite{xue2014blind}, FRIQUEE~\cite{ghadiyaram2017perceptual}, CORNIA~\cite{ye2012unsupervised}, DIIVIN~\cite{moorthy2011blind}, BLINDS-II~\cite{saad2012blind}, HOSA~\cite{xu2016blind}, Le-CNN~\cite{kang2014convolutional}, BIECON~\cite{kim2016fully}, WaDIQaM~\cite{bosse2017deep}, ResNet-ft~\cite{kim2017deep}, IW-CNN~\cite{kim2017deep}, DB-CNN~\cite{kim2017deep}, CaHDC~\cite{wu2020end}, HyperIQA~\cite{su2020blindly} and Hallucinated-IQA~\cite{lin2018hallucinated}. The comparison results are shown in Tables~\ref{tab:three} and~\ref{tab:kad}. All our experiments are conducted under ten times random train-test splitting operations, and the average SRCC and PLCC values are reported as final statistics. From the table, we can observe that all competing models achieve comparable performance on LIVE database while the performance vary on more challenging databases: TID2013 and KADID-10k. Comparing with hand-crafted based methods like BRISQUE, FRIQUEE and CORNIA, the CNN-based methods can achieve superior performance on different databases, revealing that human-perception relevant features can be learned from the training set. \bl{In particular, the PR image in Hallucinated-IQA is generated by  a pixel-level  generative network, leading to high demand for the number of training samples. This assumption can be verified by its poor performance on the CSIQ database.} Moreover, we can also find that our method achieves the best performance on CSIQA and KADID-10k  databases in terms of both SRCC and PLCC.
%Although our  method achieves the second place on LIVE database and TID2013 database, the results are still comparable  to the best model DB-CNN. Furthermore, different from DB-CNN, the external databases are not required for training by our proposed method. 

As we train our model in a paired manner, the FR results can also be acquired during the testing by involving the reference image. Herein, we also provide the FR results denoted as FPR (FR) in Tables~\ref{tab:three} and~\ref{tab:kad}. From the tables, we can observe that our FR model can achieve higher performance when compared with our NR model, as the pristine image provide more accurate reference information for quality evaluation. We also observe that the performance of our FR model is not  as good as some other FR models \textit{e.g.,} WaDIQaM-FR~\cite{bosse2017deep}. We believe this is reasonable, as the learning capability of our NR model must be considered simultaneously during the extraction of reference feature. %This constraint leads to the reference feature learning is not completely targeted at a FR task. 
%From the Table, we can observe that our method achieves the best performance on most settings and significantly outperforms them. These results reveal the effectiveness of our PR information constructed in feature level. Moreover, comparing with the method in ~\cite{hu2020tspr} where a generative adversarial network (GAN) is utilized to restore the reference information at image-level, the light network utilized in our method can significantly reduce the inference time.  
\bl{ In Table~\ref{tab:pwrc}, we futher report the performance comparison in terms of PWRC.  For a fair comparison, the same data splittings are performed for each method, and the average results of 10 repeated experiments are reported. From the Table~\ref{tab:pwrc}, we can observe that our method achieves the best performance on each dataset. A significant performance gain can be acquired by our method on the TID2013 dataset, revealing our method is also effective for ranking image pairs with higher quality levels.}

Furthermore, we compare our method with two PR image based NR-IQA methods named BPRI~\cite{min2017blind} and TSPR~\cite{hu2020tspr}.  Following the experimental setting in \cite{hu2020tspr} that four shared distortion types, \textit{i.e.}, JPEG, GB, WN and JP2K in the TID2013, LIVE and CSIQ databases are used for performance comparison. \bl{The results are presented in Table~\ref{tab:FPR}. We can observe that our method achieves the best performance in most settings and significantly outperforms the comparison methods  in terms of the average SRCC values. These results reveal the effectiveness of our PR information constructed at the feature level. It also should be noted that compared with the generative adversarial networks (GAN) utilized in~\cite{lin2018hallucinated, hu2020tspr} for reference information restoration at pixel-level, the lighter network in our method can significantly reduce the inference time.}

\subsubsection{Performance on Individual Distortion Type}
To further explore the  behaviors of our proposed method, we present the performance on individual distortion type and compare it with several competing NR-IQA models. The results of experiments performed on TID2013 database and LIVE database are shown in Table~\ref{tab:sigtid} and Table~\ref{tab:siglive}, respectively. For each database, the average SRCC values of above ten settings are reported. As shown in the Table~\ref{tab:sigtid}, we can easily observe that our method can achieve the highest accuracy on most distortion types (over 60\% subsets). By contrast, lower SRCC values are obtained on some specific distortion types, \textit{e.g.,} mean shift. The reason may lie in the challenge of PR feature hallucination due to valuable information buried by the severe distortion. It is worth noting that our method achieves significant performance improvements on some noise-relevant distortion types (\textit{e.g.,} additive Gaussian noise, masked noise) and compression-relevant distortion types (\textit{e.g.,} JPEG compression, JPEG 2000 compression). The result is consistent with the performance on LIVE database, verifying the capacity that our model possesses in restoring the PR features  from different distortion types.

\begin{table*}
  \centering
  \caption{SRCC results of individual distortion types on TID2013 database. The top two results are highlighted in boldface.}
\resizebox{512pt}{120pt}{
% Table generated by Excel2LaTeX from sheet 'bi_per-tid'
\begin{tabular}{l|ccccccc|rr}
\hline
SRCC  & BRISQUE~\cite{mittal2012no} & M3~\cite{xue2014blind} & FRIQUEE~\cite{ghadiyaram2017perceptual} & CORNIA~\cite{ye2012unsupervised} & HOSA~\cite{xu2016blind} & MEON~\cite{ma2017end} & DB-CNN~\cite{kim2017deep} & \multicolumn{1}{c}{FPR (Ours)}  \bigstrut\\
\hline
Additive Gaussian noise & 0.711 & 0.766 & 0.730  & 0.692 &  \textbf{0.833} & 0.813 & 0.790  & \textbf{0.953}      \bigstrut[t]\\
Additive noise in color components & 0.432 & 0.56  & 0.573 & 0.137 & 0.551 &  \textbf{0.722} & 0.700   &   \textbf{0.897}     &  \\
Spatially correlated noise & 0.746 & 0.782 & 0.866 & 0.741 & 0.842 &  \textbf{0.926} & 0.826 &   \textbf{0.967}     \\
Masked noise & 0.252 & 0.577 & 0.345 & 0.451 & 0.468 &  \textbf{0.728} & 0.646 &   \textbf{0.876}     \\
High frequency noise & 0.842 & 0.900   & 0.847 & 0.815 & 0.897 &  \textbf{0.911} & 0.879 &   \textbf{0.934}      \\
Impulse noise & 0.765 & 0.738 & 0.730  & 0.616 &  \textbf{0.809} &  \textbf{0.901} & 0.708 &   0.779     \\
Quantization noise & 0.662 & 0.832 & 0.764 & 0.661 & 0.815 &  \textbf{0.888} & 0.825 &  \textbf{0.920}       \\
Gaussian blur & 0.871 &  \textbf{0.896} & 0.881 & 0.850  & 0.883 &  \textbf{0.887} & 0.859 &    0.833     \\
Image denoising & 0.612 & 0.709 & 0.839 & 0.764 & 0.854 & 0.797 &  \textbf{0.865} &     \textbf{0.944}     \\
JPEG compression & 0.764 & 0.844 & 0.813 & 0.797 & 0.891 & 0.850  &  \textbf{0.894} &     \textbf{0.923}     \\
JPEG 2000 compression & 0.745 & 0.885 & 0.831 & 0.846 &  \textbf{0.919} & 0.891 & 0.916 &    \textbf{0.923}     \\
JPEG transmission errors & 0.301 & 0.375 & 0.498 & 0.694 & 0.73  & 0.746 &  \textbf{0.772} &    \textbf{0.797}     \\
JPEG 2000 transmission errors & 0.748 & 0.718 & 0.660  & 0.686 & 0.710  & 0.716 &  \textbf{0.773} &     \textbf{0.752}     \\
Non-eccentricity pattern noise & 0.269 & 0.173 & 0.076 & 0.200   & 0.242 & 0.116 &  \textbf{0.270}  &     \textbf{ 0.559}    \\
Local bock-wise distortions & 0.207 &  \textbf{0.379} & 0.032 & 0.027 & 0.268 & 0.500   &  \textbf{0.444} &    0.265     \\
Mean shift & 0.219 & 0.119 &  \textbf{0.254} &  \textbf{0.232} & 0.211 & 0.177 & -0.009 &    0.009     \\
Contrast change & -0.001 & 0.155 &  \textbf{0.585} & 0.254 & 0.362 & 0.252 & 0.548 &     \textbf{0.699}     \\
Change of color saturation & 0.003 & -0.199 & 0.589 & 0.169 & 0.045 &  \textbf{0.684} &  \textbf{0.631} &   0.409      \\
Multiplicative Gaussian noise & 0.717 & 0.738 & 0.704 & 0.593 & 0.768 &  \textbf{0.849} & 0.711 &     \textbf{0.887}     \\
Comfort noise & 0.196 & 0.353 & 0.318 & 0.617 & 0.622 & 0.406 &  \textbf{0.752} &    \textbf{0.830}      \\
Lossy compression of noisy images & 0.609 & 0.692 & 0.641 & 0.712 & 0.838 & 0.772 &  \textbf{0.860}  &     \textbf{0.982}    \\
Color quantization with dither & 0.831 &  \textbf{0.908} & 0.768 & 0.683 & 0.896 & 0.857 & 0.833 &     \textbf{0.901}     \\
Chromatic aberrations & 0.615 & 0.570  & 0.737 & 0.696 & 0.753 &  \textbf{0.779} & 0.732 &    \textbf{0.768}     \\
Sparse sampling and reconstruction & 0.807 & 0.893 & 0.891 & 0.865 &  \textbf{0.909} & 0.855 &  \textbf{0.902} &    0.887     \bigstrut[b]\\
\hline
\end{tabular}%

}
  \label{tab:sigtid}%
\end{table*}%

\begin{table}
  \centering
  \caption{Average SRCC and PLCC results of individual distortion type on LIVE database. The top two results are highlighted in boldface.}
\resizebox{240pt}{115pt}{
% Table generated by Excel2LaTeX from sheet 'bi-per-live'
\begin{tabular}{c|ccccc}
\hline
SRCC  & \multicolumn{1}{c}{JPEG} & \multicolumn{1}{c}{JP2K} & \multicolumn{1}{c}{WN} & \multicolumn{1}{c}{GB} &  \multicolumn{1}{c}{FF} \bigstrut\\
\hline
BRISQUE~\cite{mittal2012no}  & \multicolumn{1}{c}{0.965} & \multicolumn{1}{c}{0.929} & \multicolumn{1}{c}{0.982} & \multicolumn{1}{c}{\textbf{0.964}} & \multicolumn{1}{c}{0.828} \bigstrut[t]\\
M3~\cite{xue2014blind}    & \multicolumn{1}{c}{0.966} & \multicolumn{1}{c}{0.930} & \multicolumn{1}{c}{\textbf{0.986}} & \multicolumn{1}{c}{0.935} & \multicolumn{1}{c}{0.902} \\
FRIQUEE~\cite{ghadiyaram2017perceptual}  & \multicolumn{1}{c}{0.947} & \multicolumn{1}{c}{0.919} & \multicolumn{1}{c}{0.983} & \multicolumn{1}{c}{0.937} & \multicolumn{1}{c}{0.884} \\
CORNIA~\cite{ye2012unsupervised}  & \multicolumn{1}{c}{0.947} & \multicolumn{1}{c}{0.924} & \multicolumn{1}{c}{0.958} & \multicolumn{1}{c}{0.951} & \multicolumn{1}{c}{0.921} \\
HOSA~\cite{xu2016blind}  & \multicolumn{1}{c}{0.954} & \multicolumn{1}{c}{0.935} & \multicolumn{1}{c}{0.975} & \multicolumn{1}{c}{0.954} & \multicolumn{1}{c}{\textbf{0.954}} \\
dipIQ~\cite{ma2017dipiq}   & \multicolumn{1}{c}{\textbf{0.969}} & \multicolumn{1}{c}{\textbf{0.956}} & \multicolumn{1}{c}{0.975} & \multicolumn{1}{c}{0.940} & - \\
DB-CNN~\cite{kim2017deep} & \multicolumn{1}{c}{\textbf{0.972}} & \multicolumn{1}{c}{0.955} & \multicolumn{1}{c}{0.980} & \multicolumn{1}{c}{0.935} & \multicolumn{1}{c}{0.930} \bigstrut[b]\\
\hline
FPR (Ours)  & 0.962	& \textbf{0.960}	& \textbf{0.986}	& \textbf{0.959}	& \textbf{0.966}\\

\hline
PLCC  & \multicolumn{1}{c}{JPEG} & \multicolumn{1}{c}{JP2K} & \multicolumn{1}{c}{WN} & \multicolumn{1}{c}{GB} & \multicolumn{1}{c}{FF} \bigstrut\\
\hline
BRISQUE~\cite{mittal2012no}   & \multicolumn{1}{c}{\textbf{0.971}} & \multicolumn{1}{c}{0.940} & \multicolumn{1}{c}{0.989} & \multicolumn{1}{c}{0.965} & \multicolumn{1}{c}{0.894} \bigstrut[t]\\
M3~\cite{xue2014blind}    & \multicolumn{1}{c}{\textbf{0.977}} & \multicolumn{1}{c}{0.945} & \multicolumn{1}{c}{\textbf{0.992}} & \multicolumn{1}{c}{0.947} & \multicolumn{1}{c}{0.920} \\
FRIQUEE~\cite{ghadiyaram2017perceptual}  & \multicolumn{1}{c}{0.955} & \multicolumn{1}{c}{0.935} & \multicolumn{1}{c}{0.991} & \multicolumn{1}{c}{0.949} & \multicolumn{1}{c}{0.936} \\
CORNIA~\cite{ye2012unsupervised}  & \multicolumn{1}{c}{0.962} & \multicolumn{1}{c}{0.944} & \multicolumn{1}{c}{0.974} & \multicolumn{1}{c}{0.961} & \multicolumn{1}{c}{0.943} \\
HOSA~\cite{xu2016blind}   & \multicolumn{1}{c}{0.967} & \multicolumn{1}{c}{0.949} & \multicolumn{1}{c}{0.983} & \multicolumn{1}{c}{\textbf{0.967}} & \multicolumn{1}{c}{\textbf{0.967}} \\
dipIQ~\cite{ma2017dipiq}  & \multicolumn{1}{c}{0.980} & \multicolumn{1}{c}{0.964} & \multicolumn{1}{c}{0.983} & \multicolumn{1}{c}{0.948} & - \\
DB-CNN~\cite{kim2017deep} & \multicolumn{1}{c}{0.986} & \multicolumn{1}{c}{\textbf{0.967}} & \multicolumn{1}{c}{0.988} & \multicolumn{1}{c}{0.956} & \multicolumn{1}{c}{0.961} \bigstrut[b]\\
\hline
FPR (Ours)  & 0.960	& \textbf{0.971}	& \textbf{0.991}	& \textbf{0.971}	& \textbf{0.977} \bigstrut[t]\\
\hline
\end{tabular}%
}
  \label{tab:siglive}%
\end{table}%

\subsection{Cross-Database Evaluation}
To verify the generalization capability of our FPR model, we further evaluate our model on cross-database settings. We compare our method with seven NR-IQA methods, including: BRISQUE, M3, FRIQUEE, CORNIA, HOSA and two CNN-based counterparts DIQam-NR and HyperIQA. The results of DIQam-NR are reported from the original paper, and we re-train the HyperIQA by the source codes provided by the authors. All experiments are conducted with one database as training set and the other two databases as testing sets. We  present the experimental results in \ref{tab:cross}, from which we can find the model trained on LIVE (CSIQ) is  easier to generalize to CSIQ (LIVE) as  similar distortion types introduced by the two databases. However, it is a much more difficult task to generalize the model trained on CSIQ or LIVE to the TID2013 database, due to these unseen distortion types involved in TID2013 database. Despite this, we can still achieve a high SRCC in the two settings, demonstrating the superior generalization capability of our method.

\begin{table*}
  \centering
  \caption{SRCC comparison on different cross-database settings. The numbers in bold are the best results.}

% Table generated by Excel2LaTeX from sheet 'bi-cross'
\begin{tabular}{l|c|c|c|c|c|c}
\hline
Training & \multicolumn{2}{c|}{LIVE } & \multicolumn{2}{c|}{CSIQ } & \multicolumn{2}{c}{TID2013} \bigstrut\\
\hline
Testing & \multicolumn{1}{l|}{CSIQ} & \multicolumn{1}{l|}{TID2013} & \multicolumn{1}{l|}{LIVE} & \multicolumn{1}{l|}{TID2013} & \multicolumn{1}{l|}{LIVE} & \multicolumn{1}{l}{CSIQ} \bigstrut\\
\hline
BRISOUE~\cite{mittal2012no}  & \multicolumn{1}{l|}{0.562} & \multicolumn{1}{c|}{0.358} & \multicolumn{1}{c|}{0.847} & \multicolumn{1}{c|}{0.454} & \multicolumn{1}{l|}{0.790} & \multicolumn{1}{l}{0.590} \bigstrut[t]\\
M3~\cite{xue2014blind}    & \multicolumn{1}{l|}{0.621} & \multicolumn{1}{c|}{0.344} & \multicolumn{1}{l|}{0.797} & \multicolumn{1}{c|}{0.328} & \multicolumn{1}{l|}{\textbf{0.873}} & \multicolumn{1}{l}{0.605} \\
FRIOUEE~\cite{ghadiyaram2017perceptual} & \multicolumn{1}{l|}{\textbf{0.722}} & \multicolumn{1}{c|}{\textbf{0.461}} & \multicolumn{1}{l|}{0.879} & \multicolumn{1}{c|}{0.463} & \multicolumn{1}{l|}{0.755} & \multicolumn{1}{l}{ 0.635} \\
CORNIA~\cite{ye2012unsupervised}  & \multicolumn{1}{l|}{0.649} & \multicolumn{1}{c|}{0.360} & \multicolumn{1}{l|}{0.853} & \multicolumn{1}{c|}{0.312} & \multicolumn{1}{l|}{0.846} & \multicolumn{1}{l}{0.672} \\
HOSA~\cite{xu2016blind}  & \multicolumn{1}{l|}{0.594} & \multicolumn{1}{c|}{0.361} & \multicolumn{1}{l|}{0.773} & \multicolumn{1}{c|}{0.329} & \multicolumn{1}{l|}{0.846} & \multicolumn{1}{l}{0.612} \\
DIQaM-NR~\cite{bosse2017deep}  & \multicolumn{1}{l|}{0.681} & \multicolumn{1}{c|}{0.392} &    -   &    -   &     -  & \multicolumn{1}{l}{\textbf{0.717}} \\
HyperIQA~\cite{wu2020end}  &   \textbf{0.697}    &  \textbf{0.538}     &   \textbf{0.905}    &    \textbf{0.554}   &   0.839    & 0.543 \bigstrut\\
\hline
FPR (Ours)  &    0.620   &   0.433    &   \textbf{0.895}    &  \textbf{0.522}     &   \textbf{0.884}    & \textbf{0.732}  \bigstrut\\
\hline
\end{tabular}%

  \label{tab:cross}%
\end{table*}%

\subsection{Ablation Study}
In this subsection, to reveal the functionalities of different modules in our proposed method, we perform the ablation study on TID2013 database. To be consistent with the experimental setting on intra-database,  60\%, 20\%, 20\% images in TID2013 are grouped for training, validation and testing sets without content overlapping. Herein, we only report the ablation results by one fixed experimental splitting in Table~\ref{tab:abl}. In particular, we first ablate the PR and INN modules from our model and retain the Integrity Feature Extractor and \bl{GRU-based} Quality Aggregation modules. The \bl{performance drops dramatically} due to the fact that no extra constraint be introduced to prevent the over-fitting problem. Then we replace the INN module by directly concatenating the learned pseudo reference feature and distortion feature for quality regression, resulting in the second ablation setting. The lower SRCC (0.86 \textit{v.s} 0.89) reveals that more generalized model can be learned by our INN module. 
\bl{As described before, the triplet loss $\mathcal{L}_{trip}$ is adopted to learn more discriminative features. In this sense, we ablate the $\mathcal{L}_{trip}$ in our third experiment. Again, a  significant performance drop can be observed. The reason may lie in the quality discriminative feature learning resulting from the constraint of $L_{trip}$. Through $L_{trip}$, we build the PR feature  $F^{PR}$ and real one  $F^{R}$ by reducing their feature distance while enlarging the distance between the reference features ($F^{R}$, $F^{PR}$) and the corresponding distortion features ($F^{D}$ and $\hat{F^{D}}$). Such constraints can force the model to exhaustively explore the feature difference caused by the distortion, and finally, the quality awareness and discrimination capability can be enhanced.}
Finally, three patch score aggregation modules are compared in Table~\ref{tab:abl}. The superior performance further demonstrates the effectiveness of our \bl{GRU-based} score aggregation module.

\begin{table*}
  \centering
  \caption{SRCC performance with ablation studies performed on the TID2013 database.}
% Table generated by Excel2LaTeX from sheet 'ablate'
\begin{tabular}{c|c|c|c|c|c|c|c}
\hline
\multirow{2}[4]{*}{Exp.ID} &\multirow{2}[4]{*}{PR} & \multirow{2}[4]{*}{INN} & \multirow{2}[4]{*}{$\mathcal{L}_{trip}$} & \multicolumn{3}{c|}{Patch Aggregation} & \multirow{2}[4]{*}{SRCC} \bigstrut\\
\cline{5-7}    &   &    &   & Mean  & Weighted & GRU   &  \bigstrut\\
\hline
1 &\XSolidBrush     & \XSolidBrush   & \XSolidBrush  &       &       & \Checkmark     & 0.670 \bigstrut[t]\\

2 &\Checkmark      & \XSolidBrush   & \Checkmark  &       &       & \Checkmark      &  0.859\\

3 &\Checkmark      & \Checkmark   &\XSolidBrush  &       &       & \Checkmark      &  0.772\\

4 &\Checkmark      & \Checkmark   & \Checkmark  & \Checkmark      &       &       &  0.869 \\
5 &\Checkmark      & \Checkmark  &\Checkmark    &       & \Checkmark      &       & 0.848 \\
6 &\Checkmark      & \Checkmark  & \Checkmark   &       &       & \Checkmark      & 0.887 \bigstrut[b]\\
\hline
\end{tabular}%
  \label{tab:abl}%
\end{table*}%
%----------------------------------------------------------
\subsection{Feature Visualization}
\begin{figure}[t]
\begin{minipage}[b]{1.0\linewidth}
  \centerline{\includegraphics[width=1\linewidth]{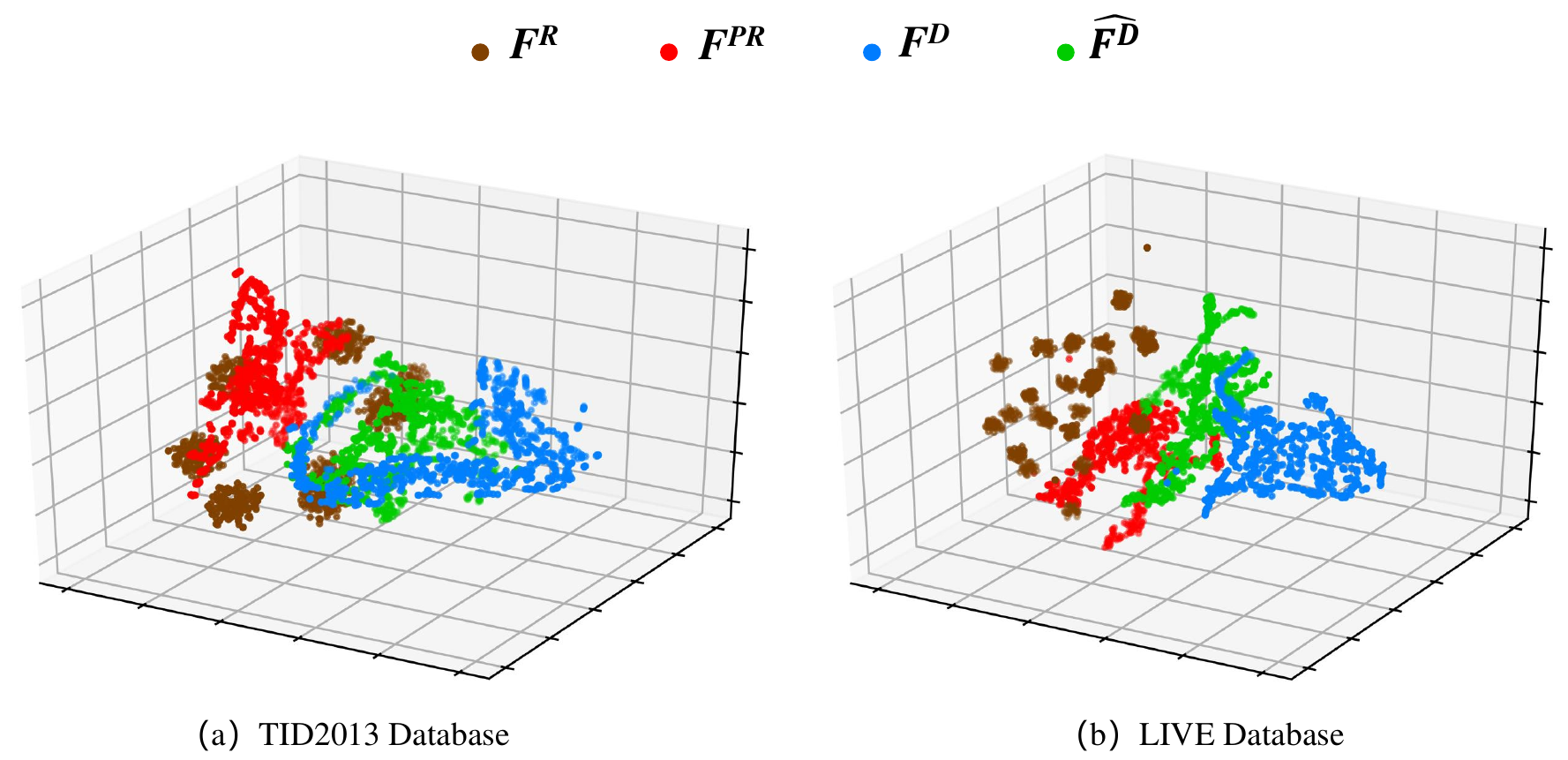}}
\end{minipage}
\caption{{T-SNE visualization of the features extracted from TID2013 and LIVE databases.  }}
\label{fig:tsne}
\end{figure}

To better understand the performance of our proposed method, we visualize the quality relevant features  $\textit{F}^{R}$, $\textit{F}^{PR}$, $\textit{F}^{D}$ and $\hat{\textit{F}^{D}}$. More specifically, we first learn two models by our method on TID2013 database and LIVE database, respectively. Then 900 image pairs of each database are randomly sampled from the two databases for testing. For each database, we reduce the  feature dimensions of  $\textit{F}^{R}$, $\textit{F}^{PR}$, $\textit{F}^{D}$ and  $\hat{\textit{F}^{D}}$ to three by T-SNE~\cite{maaten2008visualizing} and the results are visualized  in Fig.~\ref{fig:tsne}. As shown in Fig.~\ref{fig:tsne}, we can observe that the discrepancy of the reference feature $\textit{F}^{R}$ and the pseudo reference feature $\textit{F}^{PR}$ is  small due to the mutual learning strategy. By contrast, the large discrepancy can be acquired  between the  pseudo reference feature $\textit{F}^{PR}$  and distortion feature  $\hat{\textit{F}^{D}}$ as the triplet loss performed, leading to the better performance.

\begin{figure*}[ht]
\begin{center}
\includegraphics[width=0.8\textwidth]{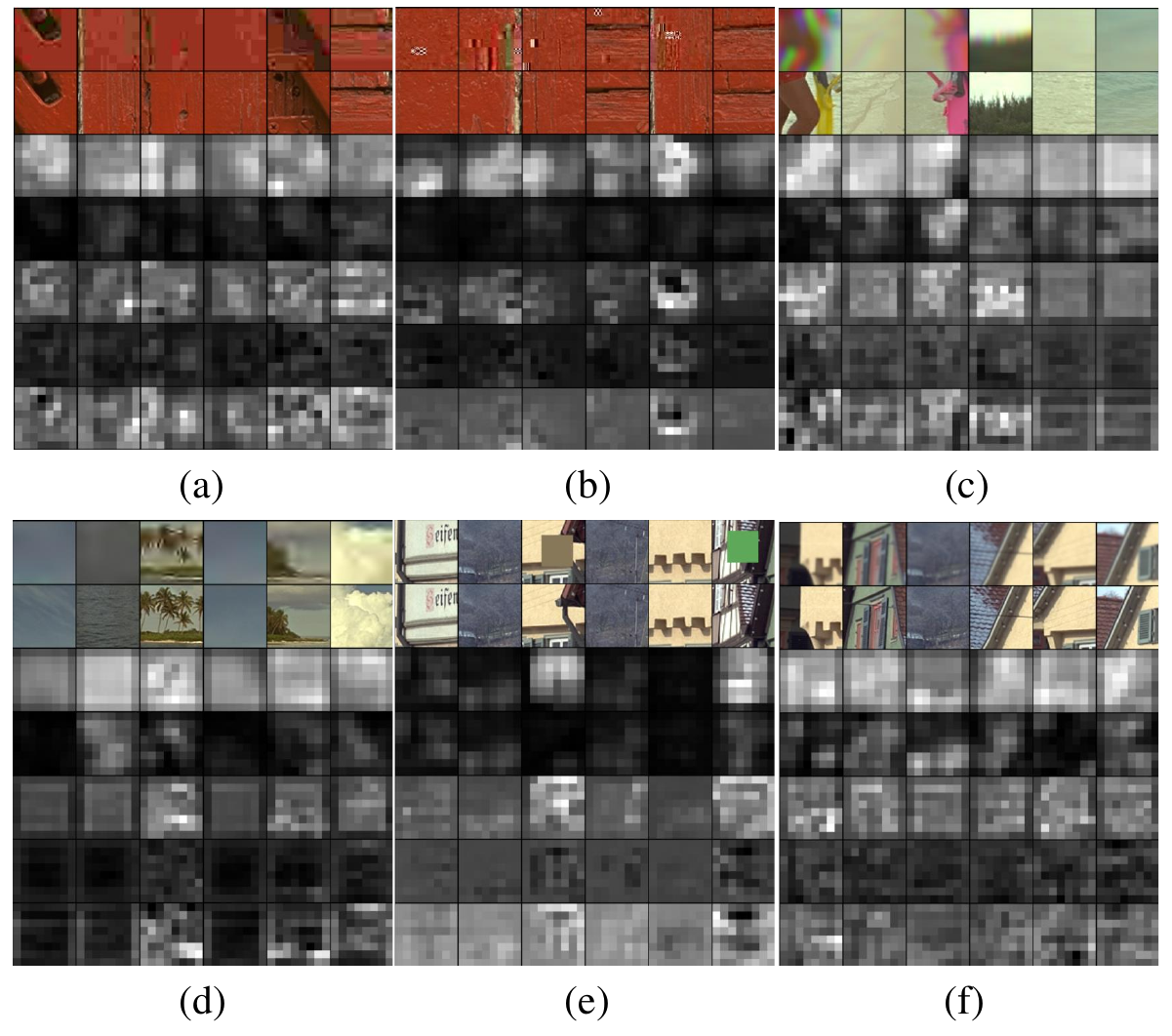}
\caption{\bl{Visualizations of the extracted feature maps. In each sub-figure, from up to down, are the sampled distorted image patches $I^R$, the reference image patches $I^D$, the corresponding  $F^D$, $F^R$, $\hat{F^{D}}$, $F^{PR}$ and $F^{NR}$ are presented, respectively. The images are distorted by: (a) JPEG compression, (b) JPEG transmission errors, (c)  Chromatic aberrations, (d)  JPEG2000 compression, (e) Local block-wise distortions of different intensity, and  (f)  Gaussian blur.}}
\label{fig:feat-show}
\end{center}
\end{figure*}

\bl{In Fig.~\ref{fig:feat-show}, we further visualize the feature maps of $F^R$, $F^D$, $F^{PR}$, $\hat{F^{D}}$ and $F^{NR}$ of several sampled distorted images.  In particular, we reduce the channel dimension of $F^D$, $F^R$, $\hat{F^{D}}$, $F^{PR}$, and $F^{NR}$ to 1 by average pooling for visualization. As shown in  Fig.~\ref{fig:feat-show}, we can observe that the distortion feature maps (${F^{D}}$ and $\hat{F^{D}}$) present a high activation on the distortion. For example, in the sub-figure (e), the local block distortion causes the high  values at the distorted block regions in ${F^{D}}$ and $\hat{F^{D}}$. The inferred $F^{PR}$ and $\hat{F^{D}}$ are consistent with $F^{R}$ and ${F^{D}}$ even for different distortion types. Meanwhile, a significant difference can be observed between $F^{PR}$ and $\hat{F^{D}}$ due to the proposed triplet loss as a constraint.}

\bl{

\subsection{Performance on Images from other Scenarios}

The quality evaluation of images from other scenarios such as  image enhancement datasets~\cite{wu2019beyond, wu2020subjective, wu2016q,min2019quality,wang2020blind} and screen content image (SCI) datasets ~\cite{yang2015perceptual,ni2017scid}  are also important tasks. To verify our model on those tasks, we further conduct experiments on one image enhancement dataset and two screen content image (SCI) datasets. In particular, the quality of the dehazing images is studied in the  SHRQ dataset~\cite{min2019quality} and the quality of the  distorted SCIs is studied in both SIQAD dataset~\cite{yang2015perceptual} and SCID dataset ~\cite{ni2017scid}. Herein, we first provide a brief overview of the above datasets as follows,

\begin{itemize}

\item SHRQ dataset \cite{min2019quality} consists of two subsets including the SHRQ-Regular and  SHRQ-Aerial. The  SHRQ-Regular includes 360  dehazed images created from 45 synthetic hazy images while the  SHRQ-Aerial includes 240  dehazed images created from 30 synthetic hazy images. In subjective testing,  subjects need to rate the quality of the dehazed images using a five-grade continuous quality scale. Besides the dehazed image, the hazy image and the reference haze-free image are also provided.

\item SIQAD dataset \cite{yang2015perceptual} contains 20 reference SCIs and 980 distorted SCIs. The distorted images are derived from seven distortion types including Gaussian Noise (GN), Gaussian Blur (GB), Motion Blur (MB), Contrast Change (CC), JPEG, JPEG2000, and Layer Segmentation based Coding (LSC). For each distortion type, seven distortion levels are generated. 

\item SCID dataset \cite{ni2017scid} consists of 1800 distorted SCIs generated by 40 reference images. In this dataset,  nine distortion types are involved including  GN, GB, MB, CC, JPEG compression, J2K, color saturation change (CSC), high-efﬁciency video coding screen content compression (HEVC-SCC), and Color quantization with dithering (CQD). Each distortion type contains five degradation levels.  All the SCIs in SIQAD are with a resolution of 1280 × 720.
\end{itemize}

We compare our method with the existing methods on those datasets and present the comparison results  in Table~\ref{tab:shrq2} and Table~\ref{tab:sci2}. For the SHRQ dataset, our method achieves comparable performance with the latest method  LieartyIQA~\cite{li2020norm}. For SCIs, our method is able to achieve the best performance on both  the SIQAD dataset and SCID dataset, revealing the high generalization capability of our method on SCIs.}

\begin{table*}
  \centering
  \caption{\bl{Performance comparison on the dehazing dataset SHRQ ~\cite{min2019quality}. The top two results are highlighted in boldface.}}
% Table generated by Excel2LaTeX from sheet 'Sheet9'

    \begin{tabular}{l|cc|cc}
    \toprule
    \multirow{2}[4]{*}{Method} & \multicolumn{2}{c|}{SHRQ-Aerial} & \multicolumn{2}{c}{SHRQ-Regular} \\
\cmidrule{2-5}          & SRCC  & PLCC  & SRCC  & PLCC \\
    \midrule
    NIQE~\cite{mittal2012making}  & 0.401  & 0.579  & 0.414  & 0.579  \\
    BRISQUE~\cite{mittal2012no} & 0.370  & 0.510  & 0.414  & 0.744  \\
    FRIQUEE~\cite{ghadiyaram2017perceptual} & 0.655  & 0.723  & 0.571  & 0.761  \\
    ILNIQE ~\cite{zhang2015feature} & 0.421  & 0.543  & 0.354  & 0.589  \\
    DIQaM-NR ~\cite{bosse2017deep} & 0.783  & 0.778  & 0.296  & 0.418  \\
    WaDIQaM-NR~\cite{bosse2017deep} & 0.864  & 0.869  & 0.636  & 0.833  \\
    HyperIQA~\cite{wu2020end}  & 0.880  & 0.881  & 0.533  & 0.767  \\
    LieartyIQA~\cite{li2020norm} & \textbf{0.914 } & \textbf{0.921 } & \textbf{0.720 } & \textbf{0.880 } \\
    \midrule
    FPR (Ours) & \textbf{0.901 } & \textbf{0.898 } & \textbf{0.692 } & \textbf{0.846 } \\
    \bottomrule
    \end{tabular}%
  \label{tab:shrq2}%
\end{table*}%

\begin{table*}[htbp]
  \centering
  \caption{\bl{Performance comparison on SIQAD \cite{yang2015perceptual} and SCID \cite{ni2017scid} datasets. The top two results are highlighted in boldface.}}
  \resizebox{500pt}{80pt}{
    \begin{tabular}{c|ccccccccc}
\toprule
    \toprule
    \multicolumn{10}{c}{SIQAD} \\ 
 
    \midrule
    Method & NIQE~\cite{mittal2012making}  & IL-NIQE~\cite{zhang2015feature} & BRISQUE~\cite{mittal2012no}  & DIIVINE ~\cite{moorthy2011blind} & QAC~\cite{xue2013learning}   & CORNIA~\cite{ye2012unsupervised} & HOSA~\cite{xu2016blind}  & BQMS~\cite{gu2016learning}  & SIQE~\cite{gu2017no} \\
    \midrule
    SRCC  & 0.359  & 0.320  & 0.775  & 0.659  & 0.301  & 0.788  & 0.718  & 0.725  & 0.763  \\
    PLCC  & 0.381  & 0.386  & 0.811  & 0.691  & 0.375  & 0.815  & 0.766  & 0.758  & 0.791  \\
    \midrule
    Method & ASIQE~\cite{gu2017no} & CLGF~\cite{wu2019blind}  & NRLT~\cite{fang2017no}  & DIQA-NR~\cite{bosse2017deep}  & WaDIQA-NR~\cite{bosse2017deep} & HRFF~\cite{zheng2019no}  & Yang~\textit{et al.} (TIP21)~\cite{yang2021no} &  Yang~\textit{et al.}(Tcy20)~\cite{yang2020no}  & FPR (Ours) \\
    \midrule
    SRCC  & 0.757  & 0.811  & 0.822  & 0.860  & \textbf{0.862}  & 0.832  & 0.834  & 0.854  & \textbf{0.873 } \\
    PLCC  & 0.788  & 0.833  & 0.844  & 0.871  & \textbf{0.877}  & 0.852  & 0.853  & 0.874  & \textbf{0.886 } \\
    \midrule
    \midrule
    \multicolumn{10}{c}{SCID} \\
  
    \midrule
    Method & NIQE~\cite{mittal2012making}  & IL-NIQE~\cite{zhang2015feature} & BRISQUE~\cite{mittal2012no}  & DIIVINE ~\cite{moorthy2011blind} & QAC~\cite{xue2013learning}   & CORNIA~\cite{ye2012unsupervised} & HOSA~\cite{xu2016blind}  & BQMS~\cite{gu2016learning}  & SIQE~\cite{gu2017no} \\
    \midrule
    SRCC  & 0.280  & 0.102  & 0.471  & 0.436  & 0.557  & 0.672  & 0.690  & 0.613  & 0.601  \\
    PLCC  & 0.322  & 0.260  & 0.520  & 0.462  & 0.585  & 0.694  & 0.711  & 0.619  & 0.634  \\
    \midrule
    Method & ASIQE~\cite{gu2017no} & CLGF~\cite{wu2019blind}  & NRLT~\cite{fang2017no}  & DIQA-NR~\cite{bosse2017deep}  & WaDIQA-NR~\cite{bosse2017deep} & HRFF~\cite{zheng2019no}  & Yang~\textit{et al.} (TIP21)~\cite{yang2021no} &  Yang~\textit{et al.}(Tcy20)~\cite{yang2020no}  & FPR (Ours) \\
    \midrule
    SRCC  & 0.605  & 0.687  & 0.609  & 0.721  & \textbf{0.758}  & -   & 0.692  & 0.756  & \textbf{0.852 } \\
    PLCC  & 0.638  & 0.698  & 0.622  & 0.740  & 0.766  & -   & 0.715  & \textbf{0.787 } & \textbf{0.856 } \\
    \bottomrule
    \end{tabular}%
  \label{tab:sci2}%
  }
\end{table*}%

%----------------------------------------------------------
\section{Conclusions}
In this paper, we propose a novel NR-IQA method named FPR by restoring the reference information at feature-level. The image quality is evaluated by measuring the discrepancy at the feature-level and the PR feature is inferred based upon the INNs. The mutual learning strategy and triplet loss ensure the   learnability and discriminability of PR features.  To aggregate the patch-wise quality scores in an image, a \bl{GRU-based} quality aggregation module is further proposed. \bl{The superior performance on the natural IQA databases, dehazing IQA databases, and screen content  IQA databases demonstrates the effectiveness of our model. Moreover, our method also achieves promising performance on the cross-dataset settings, demonstrating the high generalization capability of our model.}

%The superior performance on four synthetic databases demonstrates the effectiveness of our model.

%However, although our method can predict the image quality well on synthetic databases, the pair manner training  limits its capability of transferring the learning strategy to the authentic database. In our future work, we will explore the PR feature learning without the need of reference image by aligning their quality futures into a unify distribution. Further more, the cross-database experimental results reveal that the generalization capability of our method on unseen distortion types is still needed to be improved. This can be achieved by the domain generalization mechanisms introduced such as the domain confusion learning, domain invariant learning as well as the meta learning.

\bibliographystyle{IEEEtran}
\bibliography{NRFR}

\begin{IEEEbiography}[{\includegraphics[width=1in,height=1.25in,clip,keepaspectratio]{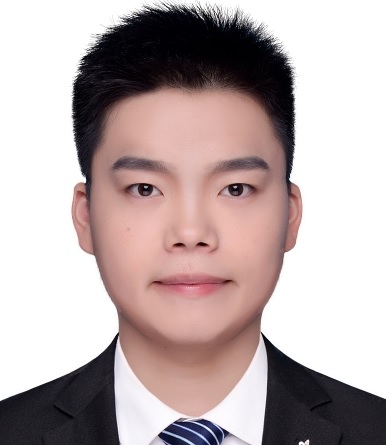}}]{Baoliang Chen} received his B.S. degree in Electronic Information Science and Technology from Hefei University of Technology, Hefei, China, in 2015, his M.S. degree in Intelligent Information Processing from Xidian University, Xian, China, in 2018, and his Ph.D. degree in computer science from the City University of Hong Kong, Hong Kong, in 2022. He is currently a postdoctoral researcher with the Department of Computer Science, City University of Hong Kong.  His research interests include image/video quality assessment and transfer learning.

\end{IEEEbiography}

\begin{IEEEbiography}[{\includegraphics[width=1in,height=1.25in,clip,keepaspectratio]{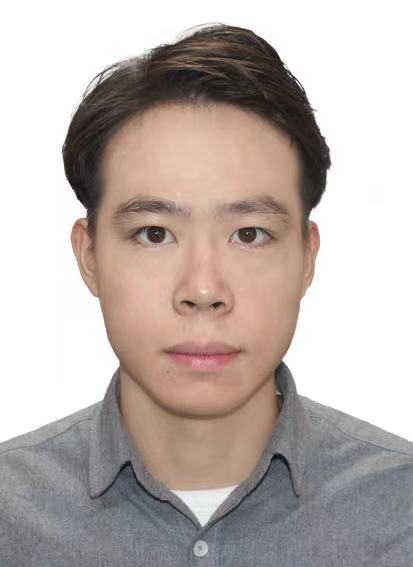}}]{Lingyu Zhu} received the B.S. degree from the Wuhan University of Technology in 2018 and the master's degree from Hong Kong University of Science and Technology in 2019. He is currently pursuing a Ph.D. degree at the City University of Hong Kong. His research interests include image/video quality assessment, image/ video processing, and deep learning.
\end{IEEEbiography}

\begin{IEEEbiography}[{\includegraphics[width=1in,height=1.25in,clip,keepaspectratio]{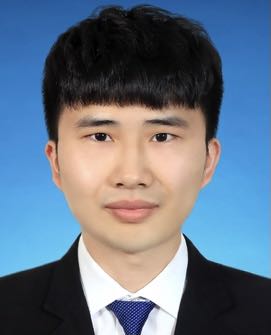}}]{Chenqi Kong} received the B.S. and M.S. degrees in Harbin Institute of Technology, Harbin, China, in 2017 and 2019, respectively. He is currently pursuing a Ph.D. degree in the Department of Computer Science, City University of Hong Kong, Hong Kong, China (Hong Kong SAR). His research interests include computer vision and multimedia forensics.
\end{IEEEbiography}

\begin{IEEEbiography}[{\includegraphics[width=1in,height=1.25in,clip,keepaspectratio]{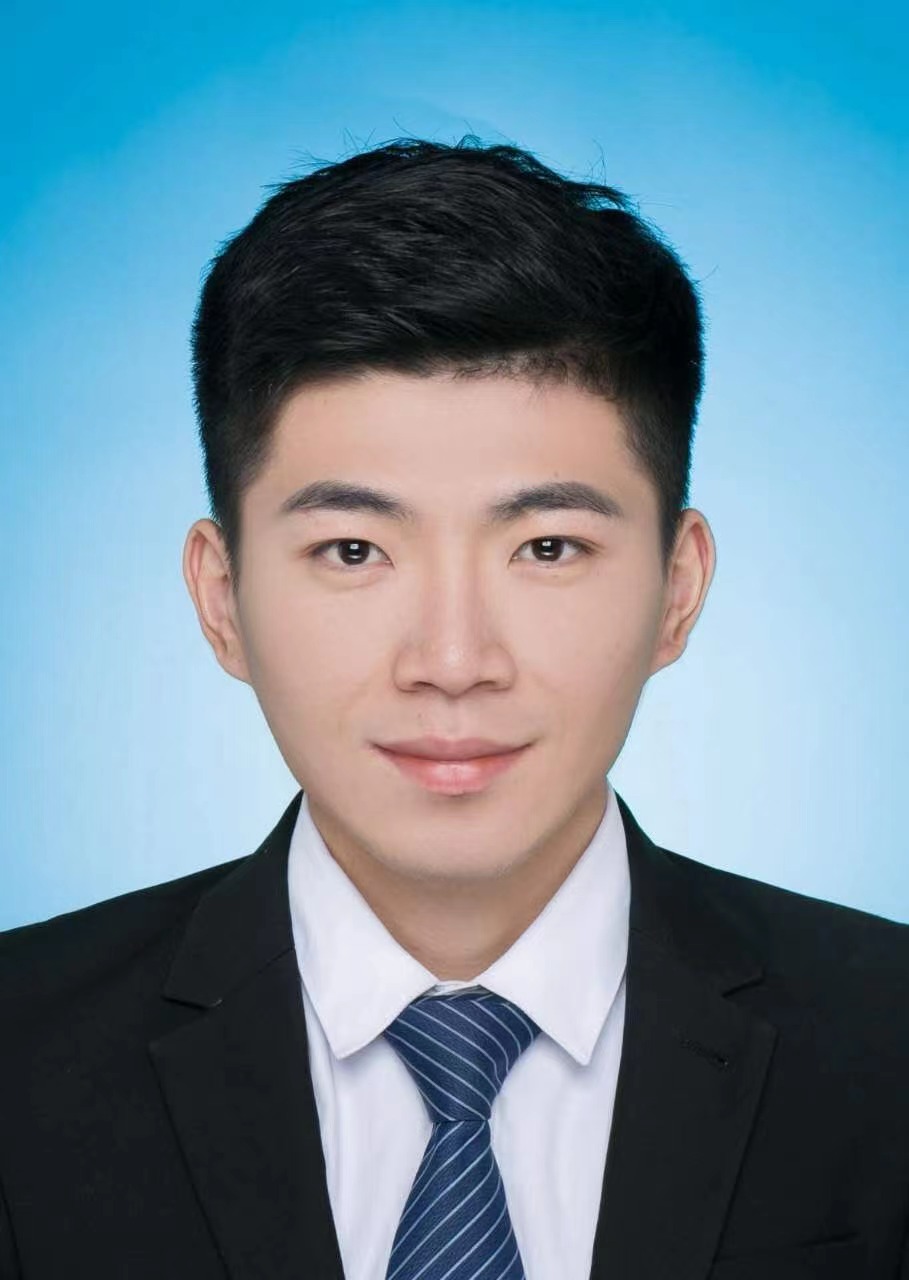}}]{Hanwei Zhu} received the B.E and M.S. degrees from the Jiangxi University of Finance and Economics, Nanchang, China, in 2017 and 2020, respectively. He is currently pursuing a Ph.D. degree in the Department of Computer Science, City University of Hong Kong. His research interest includes perceptual image processing and computational photography.
\end{IEEEbiography}

\begin{IEEEbiography}[{\includegraphics[width=1in,height=1.25in,clip,keepaspectratio]{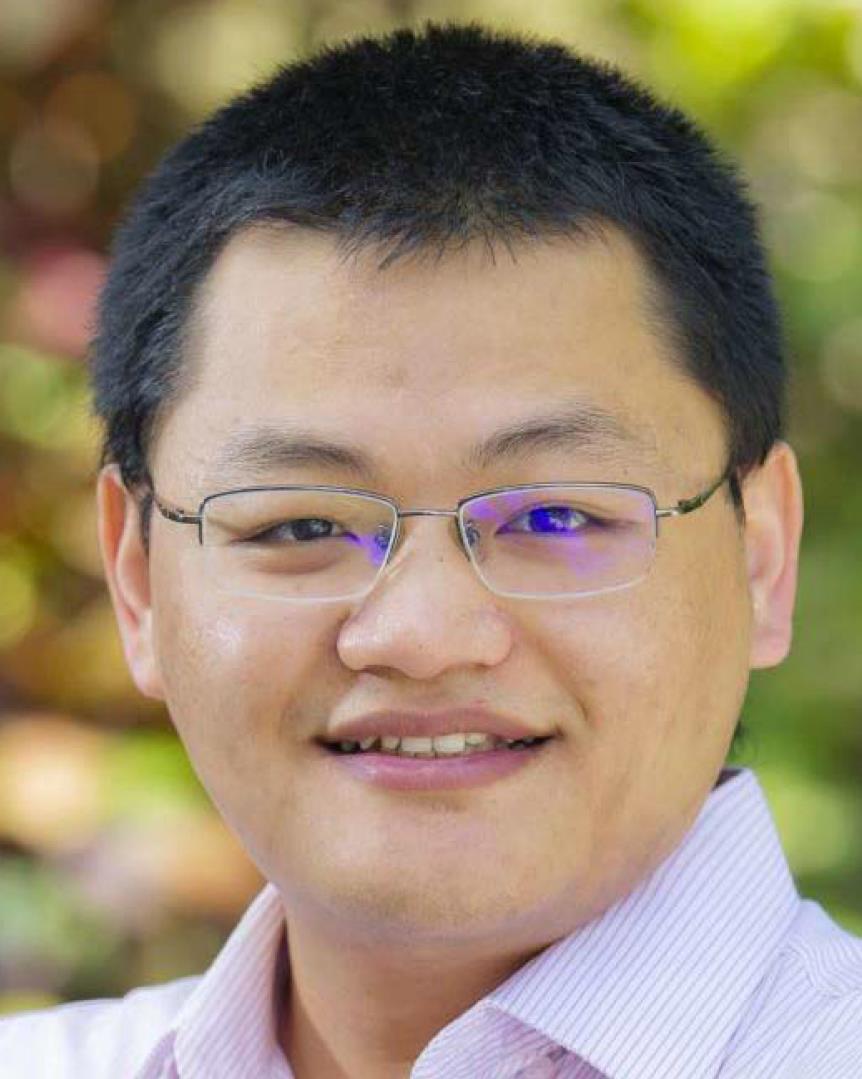}}]{Shiqi Wang}  (Senior Member, IEEE) received the B.S. degree in computer science from the Harbin Institute of Technology in 2008 and the Ph.D. degree in computer application technology from Peking University in 2014. From 2014 to 2016, he was a Post-Doctoral Fellow with the Department of Electrical and Computer Engineering, University of Waterloo, Waterloo, ON, Canada. From 2016 to 2017, he was a Research Fellow with the Rapid-Rich Object Search Laboratory, Nanyang Technological
University, Singapore. He is currently an Assistant Professor with the Department of Computer Science, City University of Hong Kong. He has proposed over 40 technical proposals to ISO/MPEG, ITU-T, and AVS standards, and authored/coauthored more than 200 refereed journal articles/conference papers. He received the Best Paper Award from IEEE VCIP 2019, ICME 2019, IEEE Multimedia 2018, and PCM 2017 and is the coauthor of an article that received the Best Student Paper Award in the IEEE ICIP 2018. His research interests include video compression, image/video quality assessment, and image/video search and analysis. 
\end{IEEEbiography}

\begin{IEEEbiography}[{\includegraphics[width=1in,height=1.25in,clip,keepaspectratio]{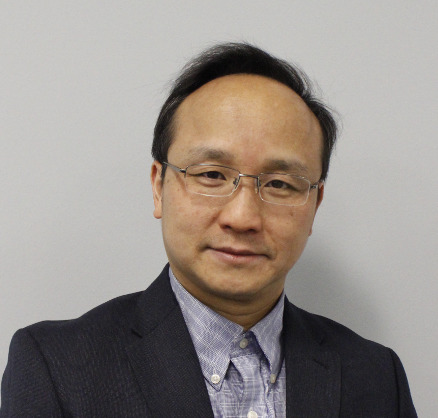}}]{Zhu Li}(Senior Member, IEEE) is a professor with the Dept of Computer Science $\&$ Electrical Engineering (CSEE), University of Missouri, Kansas City, and director of the NSF I/UCRC Center for Big Learning (CBL) at UMKC. He received his Ph.D. in Electrical $\&$ Computer Engineering from Northwestern University, Evanston in 2004. He was AFOSR SFFP summer faculty at the US Air Force Academy (USAFA), 2016-18, 2020, and 2022. He was Sr. Staff Researcher/Sr. Manager with Samsung Research America's Multimedia Standards Research Lab in Richardson, TX, 2012-2015, Sr. Staff Researcher/Media Analytics Lead with FutureWei (Huawei) Technology's Media Lab in Bridgewater, NJ, 2010~2012, an Assistant Professor with the Dept of Computing, The Hong Kong Polytechnic University from 2008 to 2010, and a Principal Staff Research Engineer with the Multimedia Research Lab (MRL), Motorola Labs, from 2000 to 2008. His research interests include point cloud and light field compression, graph signal processing and deep learning in the next gen visual compression, image processing and understanding. He has 50+ issued or pending patents, 180+ publications in book chapters, journals, and conferences in these areas. He is an IEEE senior member, associate Editor-in-Chief for IEEE Trans on Circuits $\&$ System for Video Tech, associated editor for IEEE Trans on Image Processing(2020~), IEEE Trans. on Multimedia (2015-18), IEEE Trans on Circuits $\&$ System for Video Technology(2016-19). He serves on the steering committee member of IEEE ICME (2015-18), and he is an elected member of the IEEE Multimedia Signal Processing (MMSP), IEEE Image, Video, and Multidimensional Signal Processing (IVMSP), and IEEE Visual Signal Processing $\&$ Communication (VSPC) Tech Committees. He is program co-chair for IEEE Int’l Conf on Multimedia $\&$ Expo (ICME) 2019, and co-chaired the IEEE Visual Communication $\&$ Image Processing (VCIP) 2017. He received the Best Paper Award at IEEE Int'l Conf on Multimedia $\&$ Expo (ICME), Toronto, 2006, and the Best Paper Award (DoCoMo Labs Innovative Paper) at IEEE Int'l Conf on Image Processing (ICIP), San Antonio, 2007.
\end{IEEEbiography}


%\newpage
%\clearpage
\end{document}